\documentclass[twocolumn]{aastex631}

\usepackage{booktabs}

\accepted{by AJ: August 12, 2024}

\shorttitle{JWST/NIRISS Survey for Young Brown Dwarfs}
\shortauthors{Langeveld et al.}
\begin{document}

\title{The JWST/NIRISS Deep Spectroscopic Survey for Young Brown Dwarfs and Free-Floating Planets}

\correspondingauthor{Aleks Scholz}
\email{as110@st-andrews.ac.uk}

\author[0000-0002-4451-1705]{Adam B. Langeveld}
\affiliation{Department of Physics and Astronomy, Johns Hopkins University, 3400 N. Charles Street, Baltimore, MD 21218, USA}
\affiliation{Department of Astronomy and Carl Sagan Institute, Cornell University, Ithaca, NY 14850, USA}

\author[0000-0001-8993-5053]{Aleks Scholz}
\affiliation{SUPA, School of Physics \& Astronomy, University of St Andrews, North Haugh, St Andrews, KY16 9SS, UK}

\author[0000-0002-7989-2595]{Koraljka Mu\v{z}i\'c}
\affiliation{Instituto de Astrof\'{i}sica e Ci\^{e}ncias do Espaço, Faculdade de Ci\^{e}ncias, Universidade de Lisboa, Ed. C8, Campo Grande, 1749-016 Lisbon, Portugal}

\author[0000-0001-5349-6853]{Ray Jayawardhana}
\affiliation{Department of Physics and Astronomy, Johns Hopkins University, 3400 N. Charles Street, Baltimore, MD 21218, USA}

\author[0009-0003-5245-3570]{Daniel Capela}
\affiliation{Instituto de Astrof\'{i}sica e Ci\^{e}ncias do Espaço, Faculdade de Ci\^{e}ncias, Universidade de Lisboa, Ed. C8, Campo Grande, 1749-016 Lisbon, Portugal}

\author[0000-0003-0475-9375]{Lo\"{i}c Albert}
\affiliation{Institut Trottier de recherche sur les exoplan\`{e}tes, D\'{e}partement de physique, Universit\'{e} de Montr\'{e}al, Canada}
\affiliation{Observatoire du Mont-M\'{e}gantic, Universit\'{e} de Montr\'{e}al, C.P. 6128, Succ. Centre-ville, Montr\'{e}al, H3C 3J7, Qu\'{e}bec, Canada}

\author[0000-0001-5485-4675]{Ren\'{e} Doyon}
\affiliation{Institut Trottier de recherche sur les exoplan\`{e}tes, D\'{e}partement de physique, Universit\'{e} de Montr\'{e}al, Canada}
\affiliation{Observatoire du Mont-M\'{e}gantic, Universit\'{e} de Montr\'{e}al, C.P. 6128, Succ. Centre-ville, Montr\'{e}al, H3C 3J7, Qu\'{e}bec, Canada}

\author[0000-0001-6362-0571]{Laura Flagg}
\affiliation{Department of Physics and Astronomy, Johns Hopkins University, 3400 N. Charles Street, Baltimore, MD 21218, USA}
% \affiliation{Department of Astronomy and Carl Sagan Institute, Cornell University, Ithaca, NY 14850, USA}

\author[0000-0003-1863-4960]{Matthew de Furio}
\affiliation{Department of Astronomy, University of Texas at Austin, Austin, TX 78712, USA}

\author[0000-0002-6773-459X]{Doug Johnstone}
\affiliation{NRC Herzberg Astronomy and Astrophysics, 5071 West Saanich Rd, Victoria, BC, V9E 2E7, Canada}
\affiliation{Department of Physics and Astronomy, University of Victoria, Victoria, BC, V8P 5C2, Canada}

\author[0000-0002-6780-4252]{David Lafr\`{e}niere}
\affiliation{Institut Trottier de recherche sur les exoplan\`{e}tes, D\'{e}partement de physique, Universit\'{e} de Montr\'{e}al, Canada}

\author[0000-0003-1227-3084]{Michael Meyer}
\affiliation{Department of Astronomy, University of Michigan, Ann Arbor, MI 48109, USA}

\begin{abstract}
The discovery and characterization of free-floating planetary-mass objects (FFPMOs) is fundamental to our understanding of star and planet formation. Here we report results from an extremely deep spectroscopic survey of the young star cluster NGC1333 using NIRISS WFSS on the James Webb Space Telescope. The survey is photometrically complete to $K\sim 21$, and includes useful spectra for objects as faint as $K\sim 20.5$. The observations cover 19 known brown dwarfs, for most of which we confirm spectral types using NIRISS spectra. We discover six new candidates with L-dwarf spectral types that are plausible planetary-mass members of NGC1333, with estimated masses between \mbox{5--15$\,M_{\mathrm{Jup}}$}. One, at $\sim5\,M_{\mathrm{Jup}}$, shows clear infrared excess emission and is a good candidate to be the lowest mass object known to have a disk. We do not find any objects later than mid-L spectral type ($M\lesssim 4\,M_{\mathrm{Jup}}$). The paucity of Jupiter-mass objects, despite the survey's unprecedented sensitivity, suggests that our observations reach the lowest mass objects formed like stars in NGC1333. Our findings put the fraction of FFPMOs in NGC1333 at $\sim10$\,\% of the number of cluster members, significantly more than expected from the typical log-normal stellar mass function. We also search for wide binaries in our images and report a young brown dwarf with a planetary-mass companion.
\end{abstract}

\keywords{}

\section{Introduction} 
\label{sec:intro}

The outcome of star formation is a strong function of stellar mass, with sub-solar-mass stars far outnumbering those more massive than the Sun. It is well established that for a wide range of Galactic star-forming environments, the stellar mass function is universal and can be described as a series of power laws \citep{kroupa2001} or a log-normal function \citep{chabrier2003}, without clear evidence for environmental variations \citep{bastian2010,damian2021}. It is also the current consensus that the mass function for brown dwarfs with masses down to the deuterium burning limit ($\approx 0.015\,M_{\odot}$, equivalent to 15$\,M_{\mathrm{Jup}}$) is similar across the environments investigated so far, with about 2--5 brown dwarfs formed for every 10 stars \citep{andersen2008,muzic2017,almendros2022,kirkpatrick2024}. How to explain the evidence for a universal stellar mass function is still a matter for debate.

The observational picture is much less clear for masses below the threshold for fusion processes. This domain is particularly interesting, since it is where we expect to find not only the lowest mass objects that formed like stars, but also the most massive objects that formed like planets and were subsequently ejected from their natal systems. The number of objects produced by star forming processes is expected to decline in this ultra-low-mass domain, with a hard boundary at the opacity limit for fragmentation, which has been predicted by theory but not yet been observed \citep{bate2012}. Deep surveys with the James Webb Space Telescope (JWST) offer the prospect of probing the least massive products of the star formation process. 

Meanwhile, numerical simulations predict significant numbers of ejected giant planets with masses between \mbox{1--15$\,M_{\mathrm{Jup}}$}, about 1--5\,\% of the stellar and substellar population of a young cluster -- for a detailed review and analysis of these expectations, tailored to ongoing JWST programs, we refer to \citet{scholz2022}. There are good reasons to expect that the frequency of these fusion-less, rogue objects that formed like planets would depend on environmental factors, such as the density of the birth cluster, which should affect the number of planetary ejections through stellar encounters \citep{parker2012}. 

Prior to JWST observations, the population of young stellar clusters and associations was well characterized only down to masses of $\approx$ 0.01\,$M_{\odot}$ (or 10\,$M_{\mathrm{Jup}}$). In some regions, objects with estimated masses below this limit have been found, but typically the surveys are incomplete in this mass domain and spectroscopic verification is challenging from the ground. In all regions studied to this depth from the ground or with the Hubble Space Telescope, a population of free-floating planetary-mass objects (FFPMOs) has been identified, defined here as objects with masses between 1--15\,$M_{\mathrm{Jup}}$, no matter the formation process, and not in orbit around a star. For most nearby star-forming regions, the samples hitherto are small. Two exceptions are the Orion Nebula Cluster \citep{robberto2020,gennaro2020} and Upper Scorpius \citep{lodieu2021,miretroig2022}, although spectroscopy is lacking for the full samples of these studies. JWST is currently advancing this field in two distinct ways: (1) it allows us to push the mass limit to $\approx$~1\,$M_{\mathrm{Jup}}$ with its exceptional sensitivity in the infrared; and (2) it facilitates detailed characterization of FFPMOs using its suite of instruments.

The first results from deep surveys of nearby star-forming regions with JWST were recently published in \citet{pearson2023} and \citet{luhman2024}. A few projects with similar goals are underway. Only through multi-faceted studies of diverse regions can the questions posed above be addressed. In this paper, we present initial results from a deep survey of the young star cluster NGC1333, conducted using the wide field slitless spectroscopy (WFSS) mode of the Near-InfraRed Imager and Slitless Spectrograph (NIRISS) on the JWST. In Section~\ref{sec:obs} we present the survey design and goals. In Section~\ref{sec:data} we describe the path from raw data to science-ready spectra. We then identify a sample of new planetary-mass candidate objects in NGC1333 and discuss their properties in Section~\ref{sec:newbds}. In Section~\ref{sec:binaries} we present a search for wide binaries in our field. Finally, in Section~\ref{sec:discussion} we place our result in context, then summarize our work in Section~\ref{sec:summary}.

\section{JWST/NIRISS observations}
\label{sec:obs}

The observations discussed in this paper were obtained as part of the NIRISS Guaranteed Time Observations, in program 1202 (PI: A. Scholz). We used the WFSS mode to survey a large portion of the young cluster NGC1333. WFSS is a slitless spectrograph, with a field of view of $2.2'\times2.2'$. It produces low-resolution ($R=150$) spectra over the near-infrared spectral range with a choice of filters. The observations were conducted on 2023 August 25--26. All the JWST data used in this paper were obtained from the Mikulski Archive for Space Telescopes (MAST) at the Space Telescope Science Institute. The specific observations analyzed can be accessed via \dataset[10.17909/y6ph-d557]{http://dx.doi.org/10.17909/y6ph-d557}.

\citet{doyon2023} provide an overview of the NIRISS instrument, while \citet{willott2022} describe the WFSS mode in detail; the latter includes a brief introduction to our program. 

The NIRISS WFSS campaign is a spectroscopic survey by design. It aims to obtain a spectrum of every source in the field. In contrast, most previous studies of the brown dwarf and FFPMO populations in star forming regions were conducted with multi-band photometry, followed by spectroscopy of a color-selected sample, including the recently published work on IC348 using JWST instruments by \citet{luhman2024}. The NIRCam study of the ONC by \citet{pearson2023} is based exclusively on photometry, albeit in a wide range of bands. Slitless spectroscopy is a rarely used method for this specific endeavor, but it offers an opportunity to avoid possible selection biases introduced by a color cut. One downside of this technique is the inflicted noise penalty, because the background is not suppressed by the slit \citep{willott2022}.

\subsection{Target field}
\label{sec:target_field}

\begin{figure*}[p]
   \centering
   \includegraphics[width=0.95\textwidth]{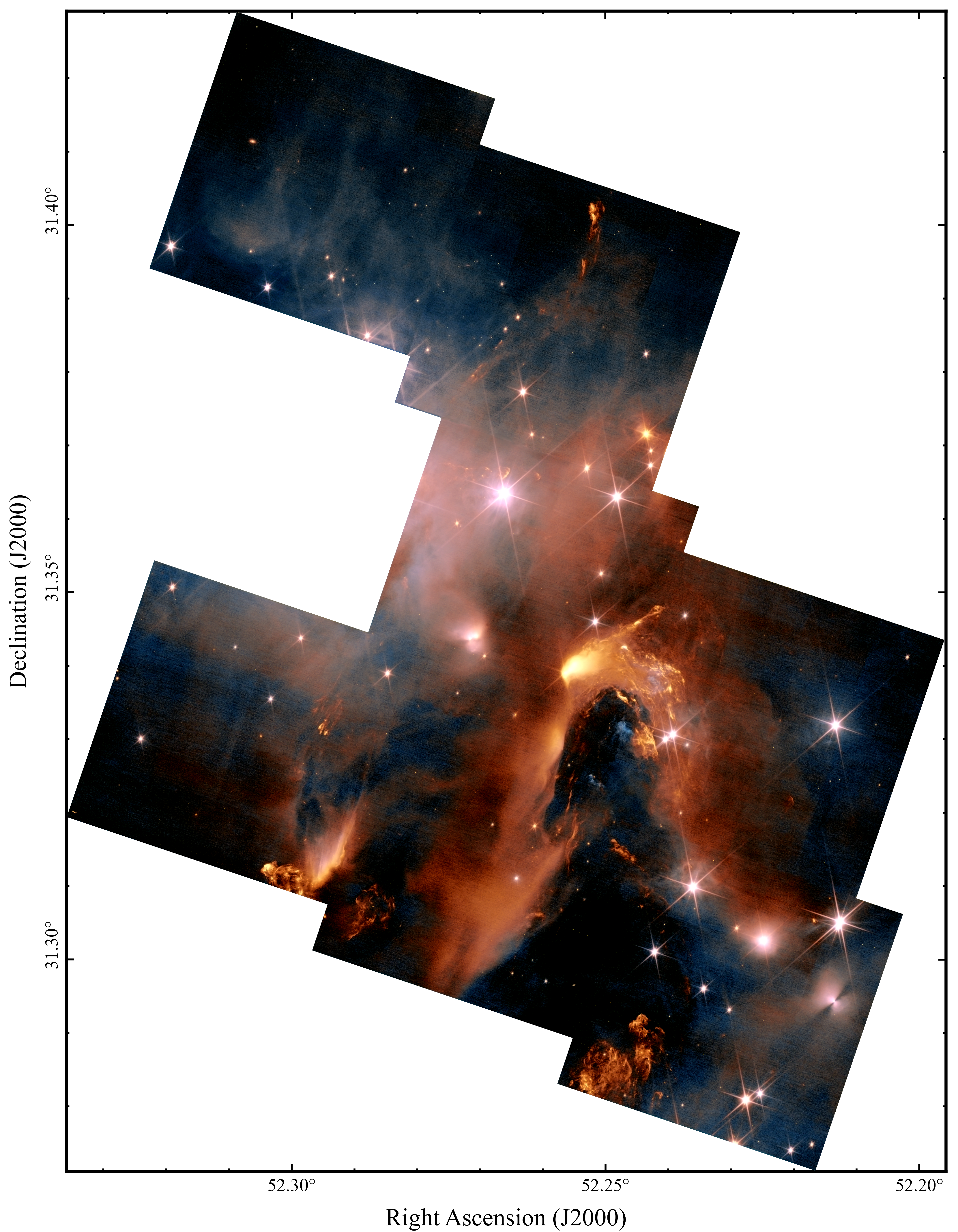}
   \caption{Color composite of the two NIRISS WFSS images of NGC1333 observed in this spectroscopic survey. To create the composite, we used the F150W image for the blue channel and the F200W image for the green and red channels. For the red/green channels, the highest/lowest values were cropped. The narrow gaps between the images in the mosaic were interpolated.}
   \label{fig:field}
\end{figure*}

NGC1333 is a compact young cluster at a distance of only $\sim 300$\,pc \citep{kuhn2019}. The cluster exhibits only moderate extinction and is 20$^{\circ}$ off the galactic plane, which makes it an ideal target for deep pencil-beam surveys. The age of the cluster is typically cited as 1--3\,Myr \citep{scholz2013}. NGC1333 shows the hallmarks of a very young cluster, such as a high disk fraction, a high number of protostars \citep{gutermuth2008} and a high fraction of variable young stars \citep{froebrich2024}. There is no evidence for an age spread beyond the 1--5\,Myr range \citep{scholz2012b}.

Our observations cover the central portion of NGC1333 in a mosaic of seven NIRISS pointings, as shown in Figure~\ref{fig:field}, with minimal overlap between the fields. The coverage was designed as a $3\times 3$ mosaic with two fields taken out (one containing bright stars, and one with very few known cluster members). The central coordinates for the mosaic are RA~3:29:03.00 and Dec~31:21:00.0. The survey area is almost continuous, with only minimal gaps between fields. The chosen fields include about 50 of the cluster members from the census by \citet{luhman2016}, which is approximately a quarter of the full population described in their paper. The mosaic also covers 19 spectroscopically confirmed brown dwarfs collated in the catalog from \citet{luhman2016}, which is about 30\% of the known brown dwarfs in this cluster \citep{scholz2023}. These objects have late M to early L spectral types and will serve as benchmarks for our spectroscopic analysis (see Section~\ref{sec:knownbds}).

It is noteworthy that NGC1333 is part of the wider Perseus star-forming complex, with several other clusters and a distributed population of young stars. The clusters and the dispersed population overlap in proper motion and distance. The two populations are difficult to separate definitively in the outskirts of the cluster \citep{pavlidou2021}. As our observations target the very core of the cluster, this should not be a concern for our study.

\subsection{Setup and expectations}
\label{sec:instrument_setup}

For each field in the mosaic, we obtained two slitless spectroscopy exposures, one with the F150W filter, one with F200W. The two filters overlap in wavelength with the H and K bands, respectively, used commonly in ground-based studies. The total integration time for each filter was 3135\,s, split into four dithers, and 18 groups per dither. The WFSS mode offers two grisms that are oriented perpendicular to each other to reduce source overlap in crowded regions. Since our target field was not expected to be affected by significant crowding, we observed only in grism GR150C, which halves the needed exposure time.

In addition to the spectroscopy, the program also included images in the same filters taken directly before and after the spectroscopy. These images were used to identify the sources in our analysis (see Section~\ref{sec:data}). For each band, the images amount to a total exposure time of 151\,s, split into five integrations and two groups per integration, with exposures at two dithered positions. The image setup was designed to replicate or slightly exceed the depth achieved in the spectroscopy. 

The total telescope time allocated for this program was 19.9\,h, of which 12.9\,h constituted science time. With this setup, we expected to reach objects with magnitudes of $H\sim 22$ and $K\sim 21$. According to evolutionary tracks \citep{phillips2020}, this is the expected brightness of a 1--2\,$M_{\mathrm{Jup}}$ object at an age of 1\,Myr without extinction -- for details of the exposure time estimate, see \citet{willott2022}. For comparison, ground-based images of this region taken by Subaru/MOIRCS are sensitive down to $K\sim 21$, but the spectroscopic follow-up only reaches objects several magnitudes above this limit \citep{scholz2009,scholz2012b}. 

Young planetary-mass objects are expected to have L and T spectral types. For a mass of 15\,$M_{\mathrm{Jup}}$ and an age of 1--3\,Myr, current evolutionary tracks predict temperatures of 2400--2600\,K \citep{phillips2020}, corresponding to late M spectral types \citep{sanghi2023}. The same conversion gives early L spectral types for $\sim10\,M_{\mathrm{Jup}}$, and mid L for $\sim5\,M_{\mathrm{Jup}}$. T-dwarfs would correspond to masses $<4\,M_{\mathrm{Jup}}$ at this age. 

Our NIRISS WFSS setup nominally covers the 1.33--1.67 and 1.75--2.22~$\mu$m wavelength ranges. This coverage does not include the ``H-band peak'' typical of young late M- and L-dwarfs \citep[shaped by H$_2$O absorption at 1.68\,$\mu$m;][]{scholz2009}, nor the CO absorption bandhead at 2.3\,$\mu$m that is characteristic of M and L objects. For late M- to T-dwarfs, we expect a rising slope at \mbox{1.3--1.6} and 1.9--2.2\,$\mu$m, caused by H$_2$O absorption. There should also be a clear drop in flux between the long wavelength end of the F150W band and the short wavelength end of the F200W band, again resulting from H$_2$O absorption. Additional structures are expected between 1.8 and 2.0\,$\mu$m, as shown by recent JWST observations \citep{miles2023}. For T-dwarfs, we also expect CH$_4$ absorption edges at 1.58 and 2.18\,$\mu$m \citep{cushing2005}.

\section{Data reduction and spectroscopy}
\label{sec:data}

\subsection{Image processing}

Stage 2 imaging data, processed by the JWST Science Calibration Pipeline version 1.11.4, were retrieved from the Barbara A. Mikulski Archive for Space Telescopes (MAST). The main purpose of these images is to create a complete input source catalog for the spectroscopy analysis, and all processing steps are geared towards that goal.  The images were first corrected for the 1/f noise pattern using the Python script available from \url{https://github.com/chriswillott/jwst}. 

The individual images exhibit a large number of bad pixels and cosmic rays, which are difficult to eliminate by stacking, given that we had only two dithered images per field. If not cleaned, these spurious sources are mistaken for astronomical objects and vastly outnumber them. As a first step, we use the extension storing the data quality flags and replace the values of all pixels with data quality flags equal to zero with a median value of all good pixels in a $10\times10$ box around a given position. The two images corresponding to the same field were then transformed to the same pixel grid using the Python package \texttt{reproject}\footnote{\url{https://reproject.readthedocs.io/en/stable/}}, and stacked using the mean, except at the positions with large differences in the values of the pixels between the two individual frames, where we instead adopt the lower value. In the described situation, the lower pixel value typically corresponds to the background and the higher one to a cosmic ray event. Since the two images are flux calibrated, stars, galaxies, and the extended emission will not have significant excursions in brightness.

Before the combination of the two frames, we plotted the positions of stars from the Gaia DR3 catalog on top of the Stage 2 frames and visually verified that no corrections to the WCS astrometry were necessary. The typical astrometric precision in our images is $\sim 0.1$\,arcsec. Finally, the seven stacked frames were combined into a single mosaic (per filter) using the  \texttt{reproject} package. A color composite of the two final mosaics is shown in Figure~\ref{fig:field}.

\subsection{Source catalog and photometry}
\label{sec:source_catalog}

An initial source catalog was produced using {\sc Source-extractor} \citep[version 2.25,][]{bertin1996} on the F200W mosaic. One main benefit of this software is that it creates a map of the background, which is variable across the field. As basic parameters, we used a detection threshold of 3$\sigma$ of the local background noise, a minimum of 5 pixels above the threshold, a filtered background, and a $7\times 7$ pixel Gaussian convolution mask for the PSF, with a FWHM of four pixels. For the background map we used a mesh size of 50 pixels and a filter size of 10 pixels. This initial catalog contains 1198 sources. 

Next we visually checked all these objects individually and retained only those that appear compact and are plausible point sources. This filtered catalog includes 609 objects. We note that the main cause of contamination in this filtered catalog are compact features in a diffuse background, likely part of Herbig-Haro emission line objects. We err on the side of inclusivity at this stage, thus the list may include some compact features that are in fact outflow knots. For reference, when running the same object search on the F150W image, the resulting catalog contains a number similar to the filtered catalog from F200W -- most of the faint emission line features are not visible in the F150W band.

To produce a source catalog that is in the correct format for spectroscopy extraction, we also run the source extraction using the \texttt{SourceCatalogStep} task of the JWST Science Calibration Pipeline, with a signal-to-noise threshold of two, a minimum of five pixels above threshold, and a background box size of five. This catalog was cross-matched with the manually cleaned {\sc Source-Extractor} catalog, in order to eliminate non-point sources and spurious detections. During this process some duplicates (typically associated with incorrectly extracted saturated stars) and marginally detected extended objects from the initial catalog were removed. The final source catalog used for spectral extraction (see Section \ref{sec:spectral_extraction}) comprised 585 objects. 

In the process of creating this catalog, we also obtained fluxes for the F200W image, a secondary product of the image analysis. We used the FLUX\_AUTO output from {\sc Source-extractor}, which is the integrated flux within an adaptively scaled aperture that is defined by the object's light distribution. We also measured fluxes for the F150W image, using the same catalog. The images are calibrated in MJy/sr -- we converted this to Jy using the PIXAR\_SR keyword in the image header which gives the size of a pixel in steradians. For the conversion to Vega magnitudes, we used a zeropoint of 1206.0\,Jy in F150W and 767.0\,Jy in F200W\footnote{F150W and F200W zeropoints are taken from \url{https://jwst-docs.stsci.edu/jwst-near-infrared-imager-and-slitless-spectrograph/niriss-instrumentation/niriss-filters}, update Nov 17 2022}. The resulting fluxes and magnitudes are in good agreement with the isophotal fluxes provided by the JWST pipeline. 

For comparison with ground-based observations, we determined offsets between magnitudes measured in the NIRISS bands and the commonly used H and K bands. For this exercise we used near-infrared photometry from the UKIDSS-GPS survey, which includes NGC1333 \citep{lucas2008}, and we limit the comparison to objects in the bracket $17<K<19$. The average difference between the F200W magnitude and the K-band magnitude is $0.53\pm 0.06$, with no noticeable magnitude or color dependence, derived from a sample of 43 objects. For the F150W band, the difference to the H-band is $0.73\pm 0.06$ (for a smaller sample of 17 objects), again without noticable magnitude dependence. We use these offsets to shift our photometry into the standard bands for comparison with the literature and for figures in this paper, but they do not go into our spectral analysis.

The comparison with the H- and K-band magnitudes shows clearly the onset of saturation, around $K\sim17$ (or F200W~$\sim 17.5$). For objects brighter than this limit, our measured fluxes will be lower limits. This includes most of the known young brown dwarfs with mid to late M spectral types in these fields. The overwhelming majority of sources in our catalog, including the planetary-mass candidates identified in this paper, are not affected by saturation.

\begin{figure}[t]
\includegraphics[width=\columnwidth]{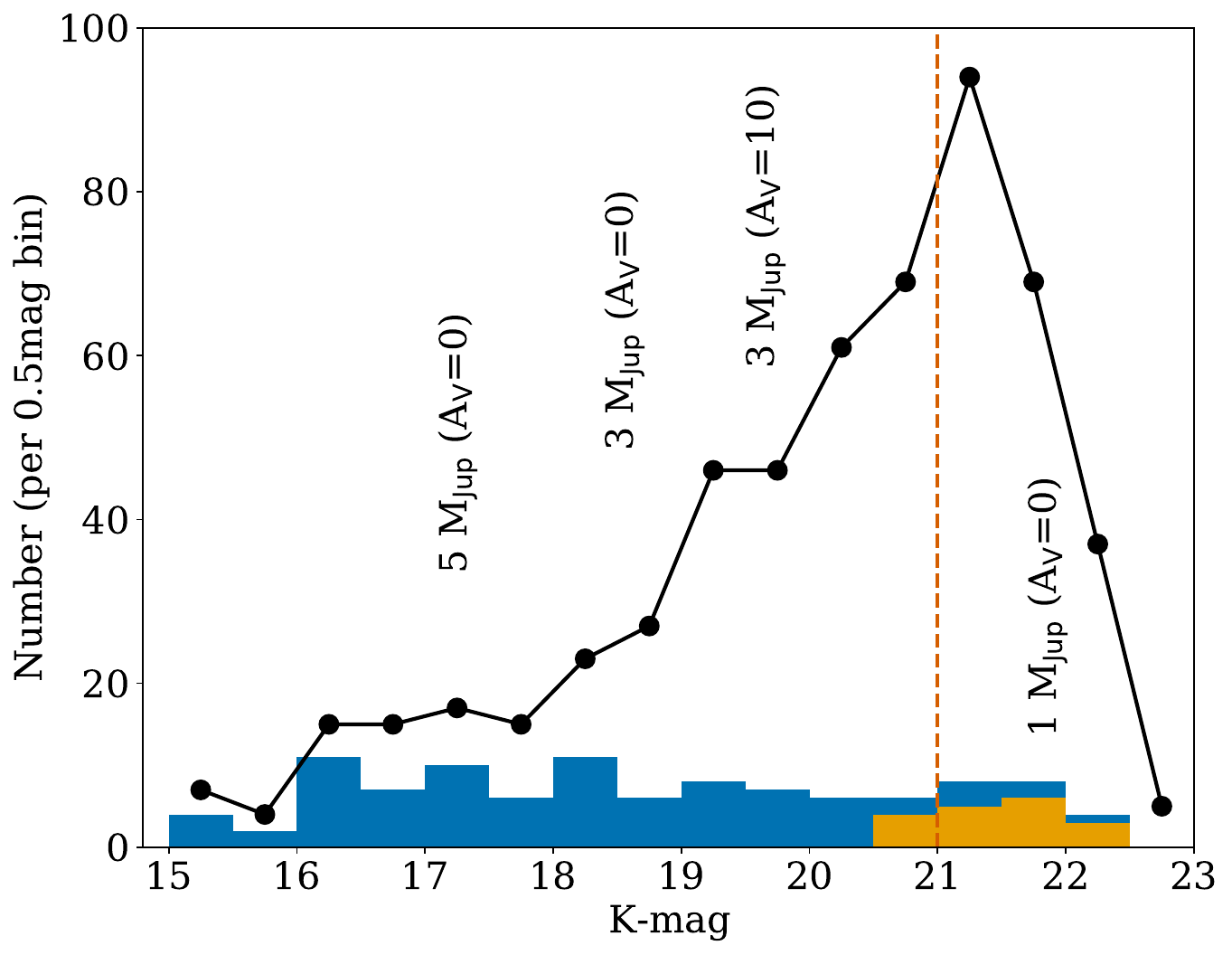} 
\caption{Histogram of the K-band magnitudes (Vega) for the catalog of objects derived from our F200W images. The by-design photometric depth of $K\sim 21$ is marked with the orange dashed line. The blue histogram is the distribution of magnitudes for our longlist candidates, examined further by spectral fitting (see Section \ref{sec:longlist}). In orange we show the distribution for objects on the longlist with very low signal-to-noise insufficient for a meaningful analysis. Some approximate mass limits are indicated, see text for details.}
\label{fig:magnitudes}
\end{figure}

Figure~\ref{fig:magnitudes} shows the distribution of K-band magnitudes for the catalog, with a bin size of 0.5\,mag. It peaks between 21.0 and 21.5, which we define as our typical photometric completeness across the field. As pointed out in Section~\ref{sec:instrument_setup}, this survey was designed to reach $K=21$. The catalog contains objects down to $K\sim 23$. To translate to masses, again using isochrones by \citet{phillips2020}: for an age of 1\,Myr, the distance of NGC1333, and no extinction, $5\,M_{\mathrm{Jup}}$ would correspond to $K=17.1$, $3\,M_{\mathrm{Jup}}$ to 18.4, and $1\,M_{\mathrm{Jup}}$ to 21.7.

\subsection{Extraction of spectra}
\label{sec:spectral_extraction}

We performed spectral extraction starting with the uncalibrated count rate (slope) images (filenames with suffix \textit{\_rate.fits}), which are the products from Stage~1 of the JWST Science Calibration Pipeline after processing the raw data. Additionally, the Stage 3 fully processed and calibrated direct images (suffix \textit{\_i2d.fits}) and segmentation maps (suffix \textit{\_segm.fits}) for the F150W and F200W filters were required. The following steps were performed in Python with version 1.12.3 of the JWST Science Calibration Pipeline \citep{bushouse2023}. The instrument-specific reference files were obtained from version 11.17.0 of the Calibration Reference Data System (CRDS).

For each of the seven NIRISS fields of NGC1333 that were observed (see Section \ref{sec:instrument_setup}), we had eight slope images -- four per filter. We performed Stage 2 corrections and calibrations to each slope image using the \texttt{run()} method to execute the \texttt{calwebb\_spec2} pipeline. Default parameters were not changed except for setting the minimum magnitude to 23. The Stage 3 processed direct image and segmentation map for the respective filter were also taken as input files, together with our custom-made source catalog (see Section~\ref{sec:source_catalog}). The same source catalog was used for both filters to ensure that spectra are extracted for the same objects, with the source object IDs and coordinates remaining consistent between the filters.

For NIRISS WFSS, the \texttt{calwebb\_spec2} pipeline performs the following steps for each slope image:
(1) transforms positions in the detector frame to a world coordinate system (WCS);
(2) subtracts a scaled background reference image from the CRDS for the respective filter and grism;
(3) divides by a flat-field reference image;
(4) extracts two-dimensional arrays from the spectral images;
(5) determines whether a spectroscopic source should be considered a point or extended object;
(6) corrects effects from overlapping spectral traces by simulating and subtracting the spectra of nearby sources (assuming a flat spectrum for each source);
(7) calibrates the flux to convert from units of count rate to surface brightness or flux density, resulting in fully calibrated (but unrectified) slope images with the filename suffix \textit{\_cal.fits};
(8) extracts a 1D signal and writes the spectra to a data product with suffix \textit{\_x1d.fits} (however, this is not used for WFSS in Stage 3). 
Bad pixels in the slope images are set to NaN and do not affect any of the following.

For a more detailed commentary on the pipeline steps, we refer to \citet{willott2022}. It is worth re-iterating that crowding is not an issue for our field, thus overlapping spectra are rare. As explained in Section~\ref{sec:longlist}, the spectra for all candidates are re-checked for contamination.

\begin{figure}[t]
    \includegraphics[width=\columnwidth]{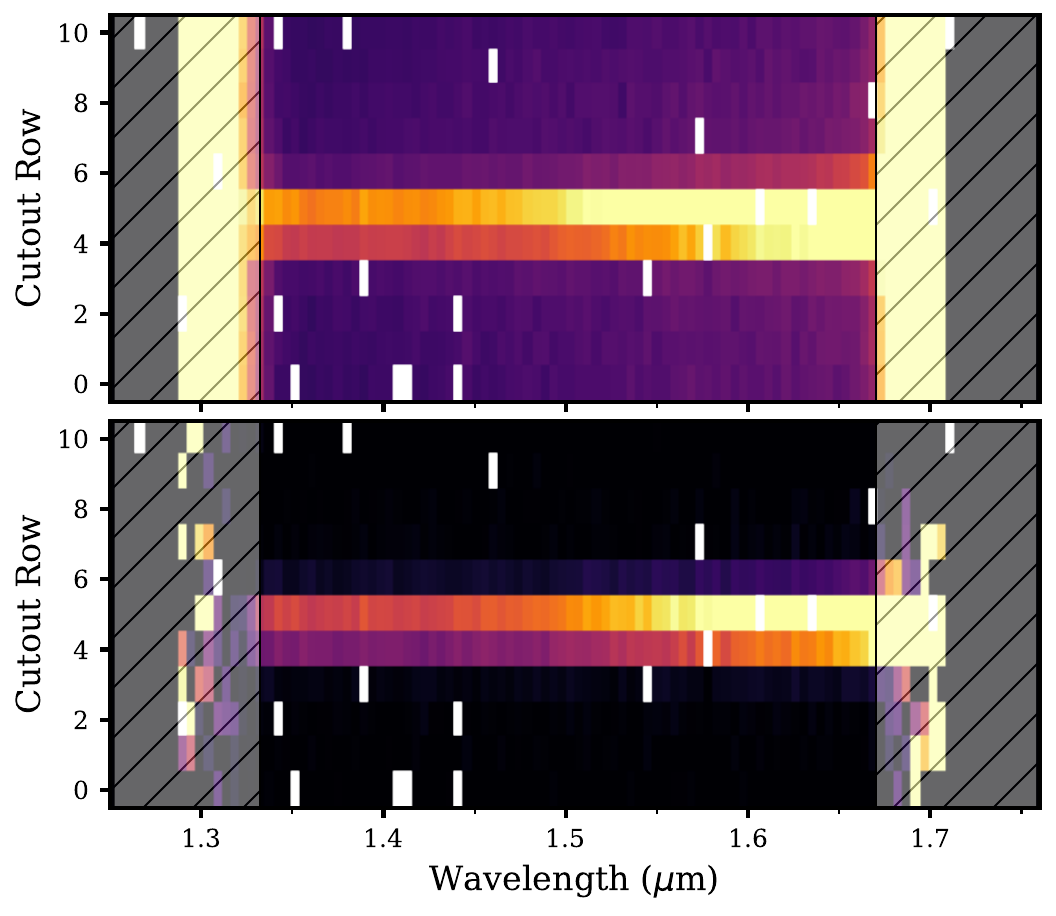}
    \caption{Example of the background subtraction process applied to the calibrated slope image cutouts between Stage~2 and Stage~3 of the JWST Science Calibration Pipeline. 
    \textit{Top panel:} Cutout for one source showing the dispersion of the spectrum (central rows). Rows above and below the spectral trail have a purple hue indicative of the background, which increases in strength towards longer wavelengths.
    \textit{Bottom panel:} Following background subtraction, the cutout now only shows significant flux from the source.
    The color scale is the same for both panels. Data inside the shaded regions are ignored when analyzing the 1D spectrum in the following analysis. White pixels are ``NaN'' values that do not affect the spectral extraction.}
    \label{fig:BGsub}
\end{figure}

Stage 2 resulted in four calibrated slope images per filter, per field. The \textit{\_cal.fits} data products contain a slope image ``cutout'' for each source in the catalog (if the spectral extraction was possible). Upon inspection, many of these cutouts appeared to contain background emission that became brighter at longer wavelengths. Therefore, we applied an extra background correction step using the Stage~2 products, before continuing with Stage~3. For each cutout, the spectral trail was identified by finding the row containing the largest flux within the middle 50\,\% of the wavelength range (effectively ignoring detector edge effects), as well as the next brightest three neighbouring rows. A column-wise median of all other rows outside of the spectral trail was calculated and subtracted from the original cutout, thus removing the background. Then, the modified cutout was saved to a new copy of the \textit{\_cal.fits} file in the same format. This process is illustrated in Figure~\ref{fig:BGsub}.

Using the modified (background-subtracted) slope image files, together with our source catalog, we executed the \texttt{calwebb\_spec3} pipeline again using the \texttt{run()} method, which combines data from multiple exposures. The result of Stage 3 is a single, co-added, one-dimensional spectrum for each object in the source catalog (providing that the spectrum could be identified in the slope images). All parameters were left at their default value. 
For NIRISS WFSS of a fixed target, the \texttt{calwebb\_spec3} pipeline performs the following steps:
(1) selects the calibrated slope image cutout for each identified source in the catalog; 
(2) extracts a one-dimensional spectrum from each of the two-dimensional cutouts; 
(3) combines the one-dimensional spectra from all exposures for the respective source using a weighted average (using the exposure times as weights), and normalizes by dividing by the sum of the weights.
For each source, the data products are named with their catalog ID and suffixes \textit{\_cal.fits}, \textit{\_x1d.fits}, and \textit{\_c1d.fits} for the three respective steps.

The spectral extraction process described above resulted in F150W and F200W spectra for 565 sources, and F150W spectra alone for an additional 20 sources. For those 20, the F200W spectrum is lost because the sources are situated close to an edge of the image, and the F200W spectral order therefore falls outside of the slope images. Many spectra show small-scale dips in the flux level which resemble absorption lines (visible for example in Figure~\ref{fig:known}). These should be considered as noise. We chose to not interpolate or remove them. As discussed in Sect. \ref{sec:instrument_setup}, the characterization of the sources relies on broad-band features, and therefore small-scale features will not matter in the following.

In the following analysis, we do not make use of the calibrated flux values from the extracted spectra. We converted the spectral density from the frequency domain (F$_{\nu}$, with units of Jy) to the wavelength domain (F$_{\lambda}$, with units of W\,m$^{-2}$\,$\micron^{-1}$) \citep{skinner1996}, such that the units are comparable to previously observed Subaru/MOIRCS spectra for the known brown dwarfs in NGC1333 \citep{scholz2009,scholz2012a,scholz2012b} -- see Section~\ref{sec:knownbds}. Spectra were then scaled to a flux of 1.0 at a wavelength of 1.40\,$\micron$ by normalizing by the average flux between 1.39 and 1.41\,$\micron$.

\subsection{Comparison with brown dwarf spectra}
\label{sec:knownbds}

\begin{figure*}[t]
    \includegraphics[width=\textwidth]{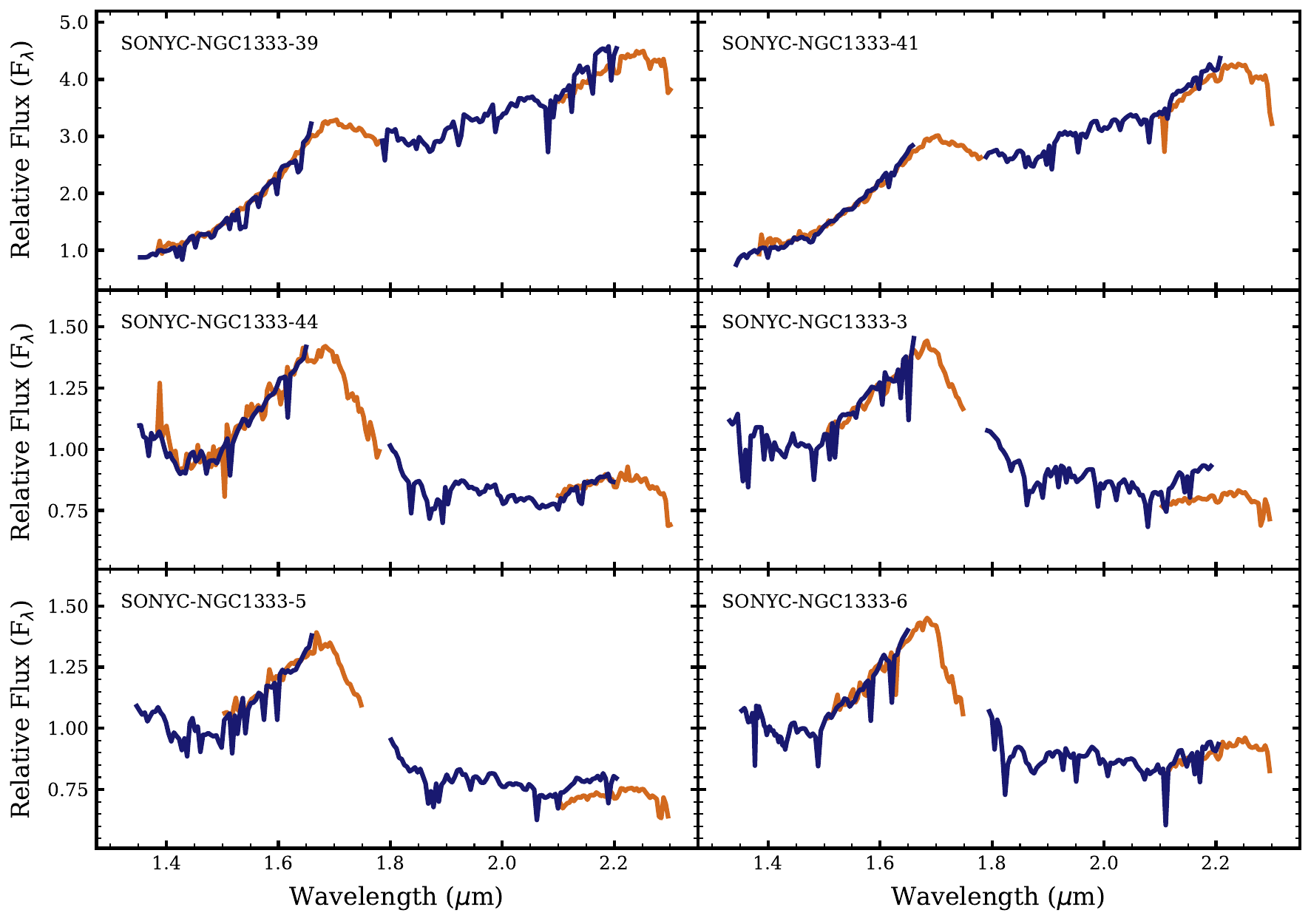}
    \vspace{-1.5em}
    \caption{Spectra of six known brown dwarfs obtained from NIRISS WFSS (dark blue), in comparison with the previously published spectra obtained with Subaru/MOIRCS (dark orange) from \citet{scholz2009,scholz2012a,scholz2012b}. NIRISS spectra were normalised to the average flux between 1.39 and 1.41\,\micron. MOIRCS spectra were scaled to match the NIRISS F150W spectra -- an offset between the NIRISS F200W data and the MOIRCS data is evident, but not consistent between objects. See Section~\ref{sec:knownbds} for further discussion.}
    \vspace{1em}
    \label{fig:known}
\end{figure*}

As described in Section~\ref{sec:target_field}, the area of this survey includes 19 spectroscopically confirmed brown dwarfs with spectral types ranging from M6 to early L. Here we use ground-based spectra from the SONYC (Substellar Objects in Nearby Young Clusters) program \citep{scholz2012a,scholz2012b} to verify the extracted NIRISS WFSS data. Figure~\ref{fig:known} shows examples of the NIRISS WFSS spectra for six of the known brown dwarfs, compared to previous observations with Subaru/MOIRCS. The literature spectra have been scaled to align with the NIRISS F150W band.

As can be appreciated from this figure, the overall shape of the NIRISS spectra follows the previously published SONYC spectra robustly. The only noticeable difference is a small offset between the NIRISS F200W band and MOIRCS K-band, which varies from object to object. For the spectra in Figure~\ref{fig:known}, as well as other objects that were compared, the offset can be cancelled out by scaling the NIRISS F200W spectra by a multiplication factor that ranges from 0.7 to 1.1, depending on the object. We attribute this factor to an unexplained calibration issue in the NIRISS data between the two grisms; the ground-based data were taken with a single grism that covers the entire H- and K-band simultaneously.

\section{Spectroscopic survey for new brown dwarfs}
\label{sec:newbds}

\subsection{Selection of candidates}
\label{sec:longlist}

Spectra for 585 sources were obtained following the spectral extraction discussed in Section~\ref{sec:spectral_extraction}. All spectra were visually checked with the goal of retaining only those with a spectral signature that could sensibly be produced by a very low mass star or brown dwarf \citep[e.g.][]{cushing2005,kirkpatrick2005,kirkpatrick2024}. In particular, we excluded objects with continuous, bright spectra, without any sign of broad molecular absorption features. We also excluded those with sudden jumps in the flux levels -- these are likely compact emission line objects. We also reject some objects with negligible flux values in the spectra. The resulting longlist comprised 114 objects. For all objects in the longlist, we visually examined the calibrated slope images (e.g. see Figure~\ref{fig:BGsub}), verifying that the spectral trace is clearly identifiable in both bands and the background is clean and flat. Parts of the spectra that are affected by residual, variable, diffuse background were masked and are not used in the following. 

In Figure~\ref{fig:magnitudes}, the distribution of magnitudes for the longlist of 114 objects is shown. It covers the full range of the magnitudes in the catalog, and in particular, it extends down to the faint limit. In orange we also show the magnitude distribution for spectra on the longlist with signal-to-noise too low for a meaningful analysis. All objects on the longlist with $K<20.5$ have spectra suitable for further analysis. Thus, the spectroscopy does not quite reach the completeness limit of the photometry, but still includes the mass range down to 1-2$\,M_{\mathrm{Jup}}$ for $A_V=0$ and 3$\,M_{\mathrm{Jup}}$ for $A_V<20$.

\subsection{Spectral fitting}
\label{sec:spectral_fitting}

To estimate spectral types (SpT), we compared the spectra of candidates on the longlist with a set of empirical templates, reddened to a range of extinctions ($A_V$). This exercise is agnostic to atmospheric physics or spectral features; it simply searches for the templates that provide the best fit. To re-iterate, the spectral fitting will follow the broad-band features, and small-scale artefacts in the spectra will not change the outcome.

Our set of empirical spectral templates is constructed as follows:
\begin{itemize}[noitemsep,nolistsep,topsep=0pt]
    \item M0 to L0 (at 0.5 subclass interval), L2, L4, L7 from \citet{luhman2017};
    \item L1 and L6 proposed by \citet[][]{allers2013} for objects 2MASS\,J05184616--2756457 and 2MASS\,J22443167+2043433 respectively;
    \item L3 constructed as an average of three objects proposed in \citet{cruz2018}, with the spectra obtained from \citet{allers2013};
    \item L5 proposed by \citet{piscarreta2024} for 2MASS\,J21543454-1055308, with the spectrum obtained from \citet{gagne2015};
    \item T0 to T8 at each 1 subclass interval from \citet{burgasser2006}.
\end{itemize}

The M-type standards have ages $<$~10 Myr, the L-types are somewhat older, belonging to nearby young moving groups, while the T-dwarf standards belong to the field. In terms of $\log{g}$ and metallicity, all templates are assumed to be similar -- subtle differences in these parameters would not be expected to have an impact on low-resolution spectra. As shown by \citet{zhang2021} and \citet{piscarreta2024}, young T-dwarfs appear slightly redder in the $J$--$K$ bands than their field counterparts; however, these differences are not relevant for low signal-to-noise spectra such as ours. 

The fitting routine is similar to that described by \citet{almendros2022} and \citet{piscarreta2024}. In short, we calculated a metric $M$ that quantifies the difference between a reddened template and the target spectrum: 
\begin{equation}
M=\frac{1}{N-3} \sum_{i=1}^{N} (O_i-T_i)^2~,
\label{eqn:metric}
\end{equation}
where $O$ is the target spectrum, $T$ is the template, and $N$ is the number of datapoints. This metric was calculated for a range of $A_V$ (0--20 with a step of 0.5\,mag). The reddening was determined following the extinction law from \citet{wang2019}, using the power law index of $-2.07$. We selected the spectral type and $A_V$ combination with minimum $M$ as the best fit. The main difference between the routine described in the aforementioned papers and the one used in this study is that we allowed the F200W spectrum to be scaled (multiplicative) up or down by a free parameter in the range of 0.7 to 1.3, in steps of 0.1, to account for the offsets described in Section~\ref{sec:knownbds}. The other difference is that the wavelength at which the spectra and the templates are normalized is kept fixed. The default normalization wavelength for the spectral fitting is 1.5\,$\mu$m if the H-band spectrum is present, and 1.9\,$\mu$m otherwise; we manually adjusted the normalization wavelength for some objects if it coincided with a noise peak.

\begin{figure*}[t]
    \includegraphics[width=\textwidth]{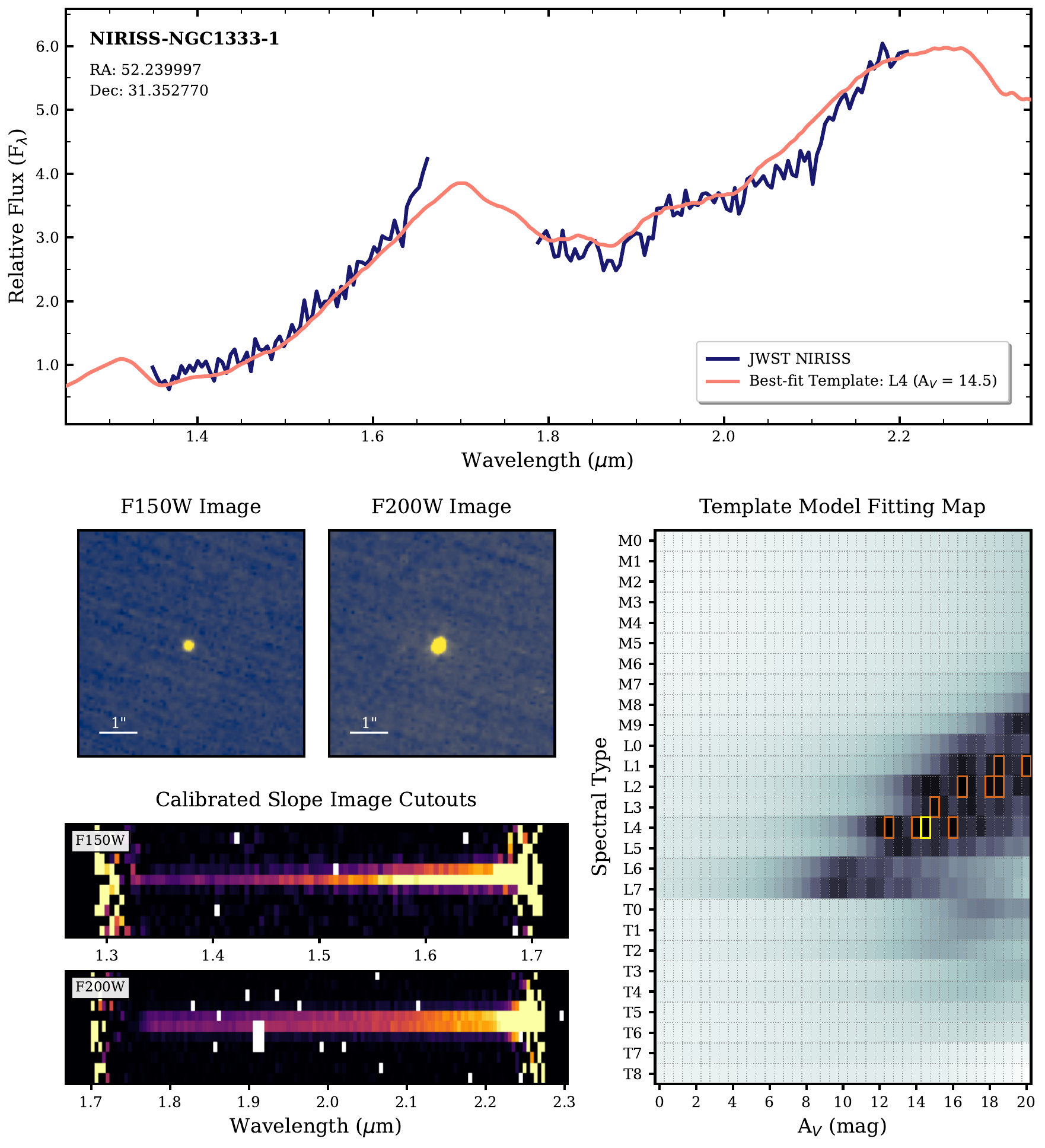}
    \caption{Results of the spectroscopic survey for object NIRISS-NGC1333-1 (NN1). 
    \textit{Top panel:} NIRISS WFSS spectrum (dark blue) with flux scaled to a value of 1 at a wavelength of 1.4\,$\micron$, and best-fit template (salmon pink) from Section~\ref{sec:spectral_fitting} that has been smoothed using a Savitzky-Golay filter.
    \textit{Middle-left panels:} NIRISS images of the object in the F150W and F200W filters with a 1\,$''$ scale.
    \textit{Bottom-left panels:} Calibrated slope image cutouts that are produced by the \texttt{calwebb\_spec3} pipeline for the NIRISS F150W and F200W filters and modified to apply the background subtraction.
    \textit{Bottom-right panel:} Map of the metric calculated as part of the spectral fitting. The templates that most closely fit the data (having lowest $M$; see equation~\ref{eqn:metric}) are located in the darkest regions. The best solution is marked with a yellow box, and the corresponding template is displayed in the top panel to compare to the observed spectrum. The remaining best nine solutions are marked in orange.}
    \label{fig:resultsNN1}
\end{figure*}

From the fitting routine, we obtain (a) a list of the best fitting spectral type and $A_V$ combinations, and (b) a visualisation of the fit results, i.e. $M$ vs. spectral type and $A_V$. A good fit is characterized by a few optimum solutions that are close together, and by a clear, well-constrained cluster of good fits in the visualisation. An example of the fitting results is shown in Figure~\ref{fig:resultsNN1}.

The known brown dwarfs in our fields give us an opportunity to vet the robustness of the spectral fitting routine. Figure~\ref{fig:sptcomp} shows the spectral type and the extinction determined by our fits against the previously published spectral type (or the average if multiple measurements exist) and extinction, taken from the census by \citet{luhman2016}, and adding one object from \citet{esplin2017}. In this figure, we plot $A_J = 0.243\,A_V$, again using the extinction law from \citet{wang2019}. Most of our spectral type and extinction estimates are within 1 subtype or 1\,mag of the literature value, with a few outliers with slightly larger deviations. In particular, our fitting routine sometimes cannot reliably distinguish between a late M spectral type with high $A_V$ and an early L type with negligible extinction, as illustrated by the two objects (blue points) where we show two of the best fitting results. Those two cases are highly reddened M8 objects, but can be equally well approximated by L-dwarf templates with little extinction. This degeneracy is expected and can only be resolved with measurements covering a wider wavelength range. Overall the comparison illustrates that our routine produces reliable results at low to moderate extinctions. 

\subsection{Newly identified planetary-mass candidates}
\label{sec:new_candidates}

\begin{figure*}[t]
\includegraphics[height=7.15cm]{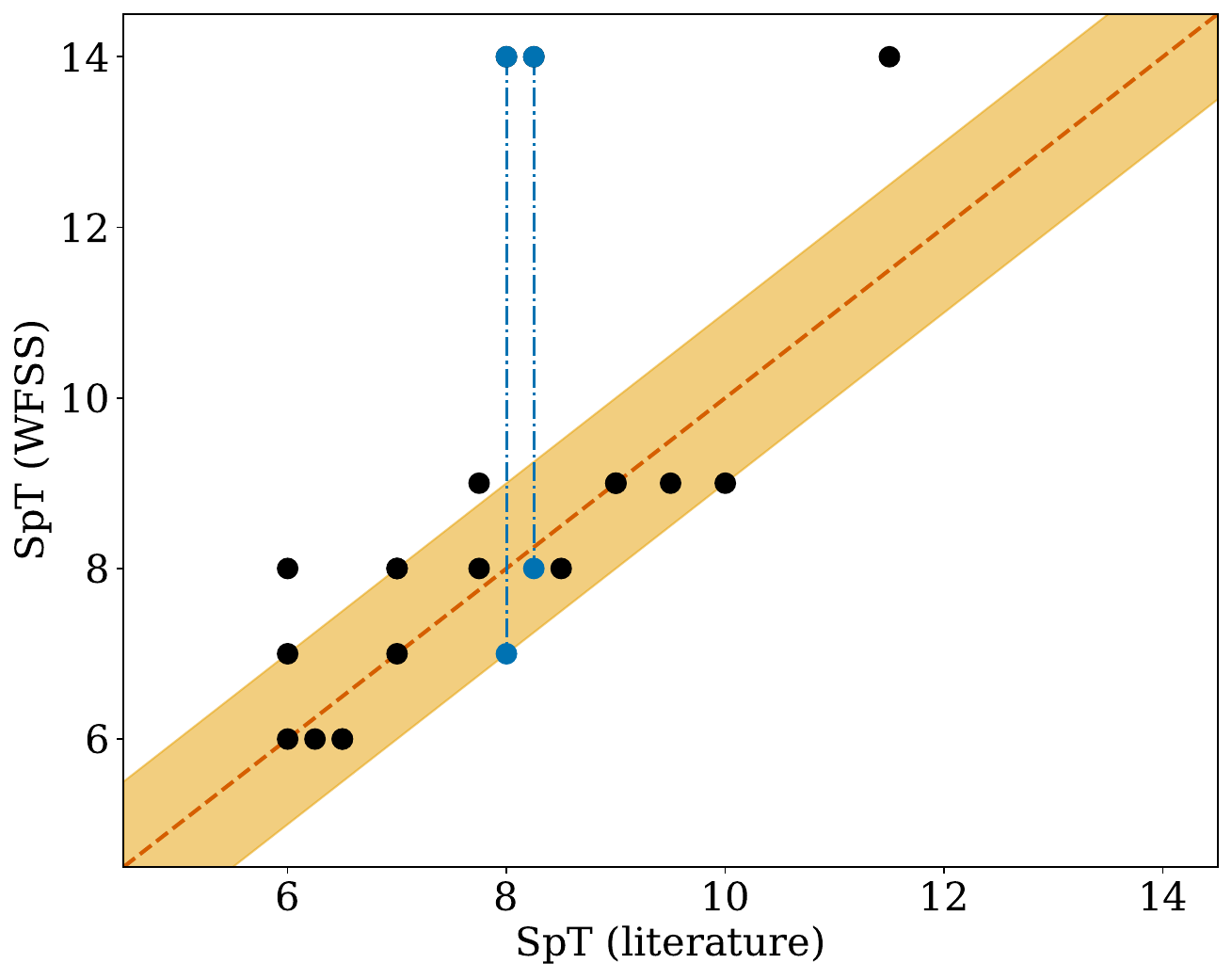} 
\hspace{0.2em}
\includegraphics[height=7.20cm]{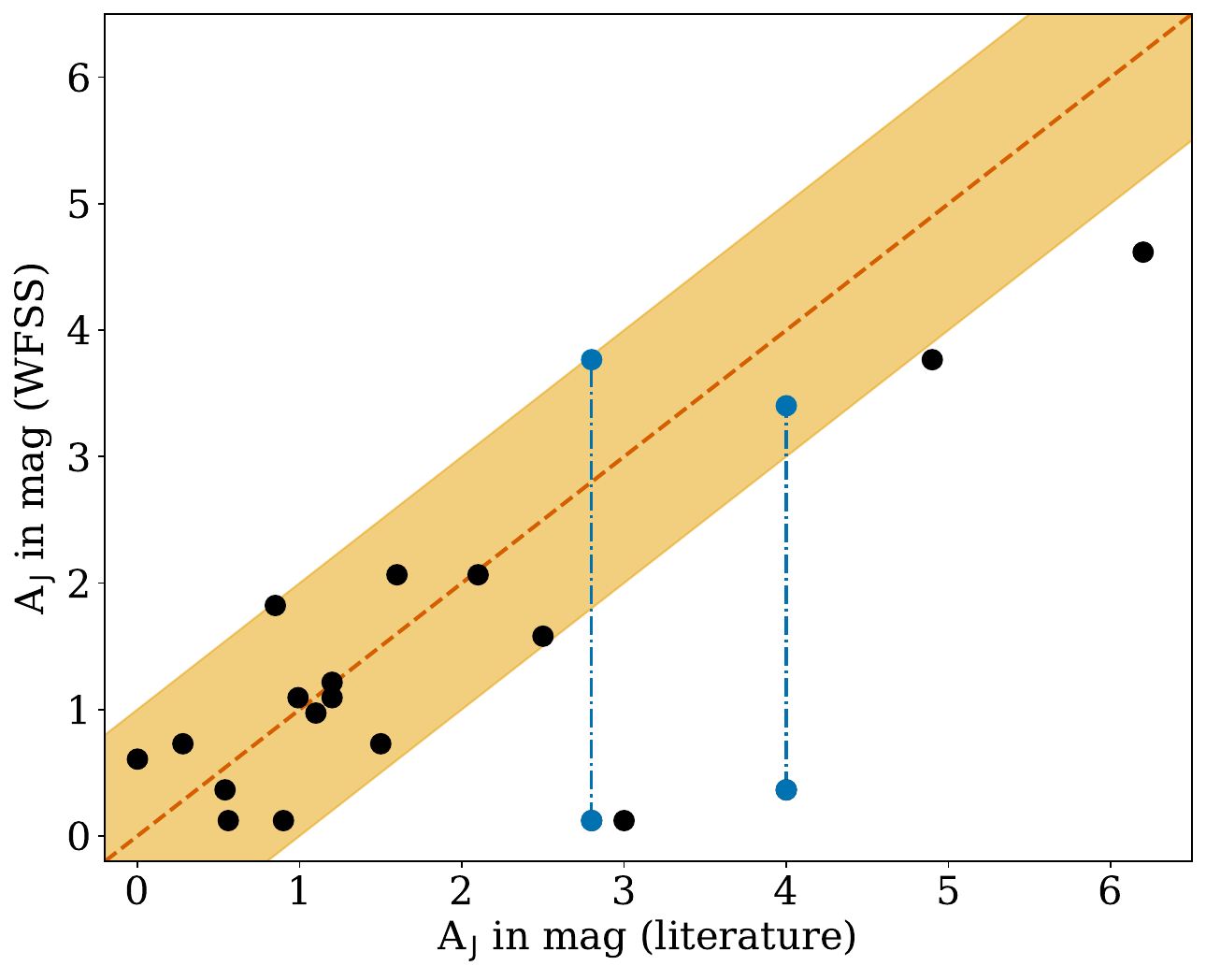} 
\caption{Comparison of the spectral type (left) and extinction $A_J$ (right) determined with our template fitting, with the average spectral type given in \citet{luhman2016}. Spectral types are coded as a linear scale, M5 is 5.0 and L2 is 12.0. The orange line marks the 1:1 relation, the shaded region corresponds to errors of 1 spectral subtype and 1\,mag in J-band extinction. For two objects (blue points) we show two possible solutions (joined with dotted-dashed lines); based on magnitudes and colors, the solution with late M spectral type is preferred.}
\vspace{0.5em}
\label{fig:sptcomp}
\end{figure*}

From the longlist, we identify 13 objects with spectral types of M9 to mid L, consistent with being planetary-mass brown dwarfs in NGC1333. Five of these have previously been spectroscopically observed, with published spectral types of M9--L1 \citep{luhman2016}, and are included in Figure \ref{fig:sptcomp}. Two more are known as M-dwarfs with infrared excess; with $K=12$ to 13, they are far too bright to be low-mass brown dwarfs. For the remaining six, we report spectroscopic classifications for the first time. The spectra for all six follow the best fitting templates. All show the expected drop-off between the two bands, the rising slope in both bands, and most also show substructure in the F200W band (see Section \ref{sec:instrument_setup}). Given their estimated spectral types, these are strong candidates to be newly discovered young free-floating planetary-mass objects in NGC1333.

Most of the other objects on the longlist have spectra consistent with being early to late M-dwarfs, among which are many known substellar members in NGC1333, as mentioned previously. Also as discussed earlier, 18 sources on the longlist have insufficient signal-to-noise for a meaningful analysis. Two more do not result in any sensible fit, i.e. no SpT/$A_V$ combination gives a plausible match. 

Notably, none of the objects on the longlist show the spectral signatures of late L- or early T-dwarfs. In particular, we do not find any objects with the typical methane absorption edges at 1.58 and 2.18$\,\mu$m. This is despite the fact that our photometric catalogue, as well as the longlist, covers objects in the magnitude range expected for T-dwarfs (see Section \ref{sec:longlist}). The magnitude limit for our spectroscopic characterization, at $K\sim 20.5$, means that we are able to detect T-dwarfs with extinctions up to $A_V\sim 20$. Despite very deep previous surveys, no cluster members with higher extinctions have been found thus far \citep{scholz2012b,luhman2016}. Thus, the lack of T-dwarfs in our spectroscopic sample points to a genuine absence or rarity of young T-dwarfs in this cluster, or alternatively, a severe overestimate of their near-infrared fluxes by evolutionary tracks.

For the six newly found planetary-mass candidates, we introduce the following naming convention: \mbox{NIRISS-NGC1333-X}, where X is replaced by a running index number, ordered by right ascension. Abbreviated, the candidates are named NN1 to NN6. The basic properties of these planetary-mass candidates are summarised in Table~\ref{tab:pmo}. We include the best fit for spectral type and $A_V$, but also an adopted spectral type given by the range of the ten best solutions, which gives an indication of the uncertainties. We also include our photometry; the candidates are faint with K-band magnitudes of 16.4 to 19.2. Most of the objects are listed by existing deep photometric surveys in NGC1333 -- in particular, most have entries in UKIDSS-GPS \citep{lucas2008}. One (NN5, see below) has several published mid-infrared measurements, in particular by \citet{rebull2015,getman2017}. None have spectra or any detailed characterization in the literature. The results from the spectral fits are visualised in Figure~\ref{fig:resultsNN1} as an example; for the remaining objects the same plots are included in Appendix~\ref{sec:appendix_figures}.

\begin{table*}[t]
\caption{List of new planetary-mass candidates in NGC1333. Spectral types (SpT) and extinction ($A_V$) are the best fits as described in Section~\ref{sec:spectral_fitting}. The adopted spectral type accounts for the ten best fitting solutions. The H- and K-band photometry comes from our own images, except for the saturated NN5 where we take it from UKIDSS-GPS \citep{lucas2008}. Typical uncertainties in the photometry are $\pm0.05$\,mag. The candidates are named NIRISS-NGC1333-X, where X is replaced by a running index number (ordered by increasing RA), or abbreviated NN1 to NN6.}
\label{tab:pmo}
\begin{tabular*}{\textwidth}{c@{\extracolsep{\fill}}ccccccccc}
\toprule
no & RA        & Dec      & SpT         & Av                   & SpT       & F150W & F200W & H   & K   \\
   & $^\circ$  & $^\circ$ & (best fit)  & (best fit)           & (adopted) & mag   & mag   & mag & mag \\
\midrule
NN1  & 52.239997 & 31.352770 & L4 & 14.5 & L1--L4 & 21.39 & 19.10 & 20.66 & 18.57 \\ %382
NN2  & 52.245580 & 31.290763 & L1 &  0.0 & M9--L1 & 21.12 & 19.74 & 20.39 & 19.20 \\ %67
NN3  & 52.273062 & 31.358947 & M9 &  6.0 & M8--L2 & 18.87 & 17.78 & 18.14 & 17.25 \\ %393
NN4  & 52.273364 & 31.394256 & L3 &  4.0 & M9--L4 & 21.22 & 19.66 & 20.49 & 19.13 \\ %482
%NN5  & 52.275820 & 31.390840 & M9 &  3.5 & M5--M9 & 21.00 & 19.76 & 20.27 & 19.23 \\ %463
NN5  & 52.280303 & 31.402649 & L4 & 11.0 & L2--L7 & 19.57 & sat.  & 18.73 & 16.37 \\ %519
NN6  & 52.309692 & 31.406801 & L4 &  7.5 & L2--L5 & 21.18  & 19.53 & 20.45 & 19.00 \\ %586
\bottomrule
\end{tabular*}
\vspace{1em}
\end{table*}
%451 is [LAL96] 231, excluded by SONYC spectrum, early-mid M dwarf. Spectrum deviates at short end.
%322 is [LAL96] 245, early-mid M dwarf, disk. Incomplete K-band coverage.
%20 is known, M9.

\begin{figure}[t]
\vspace{0.75em}
\includegraphics[width=\columnwidth]{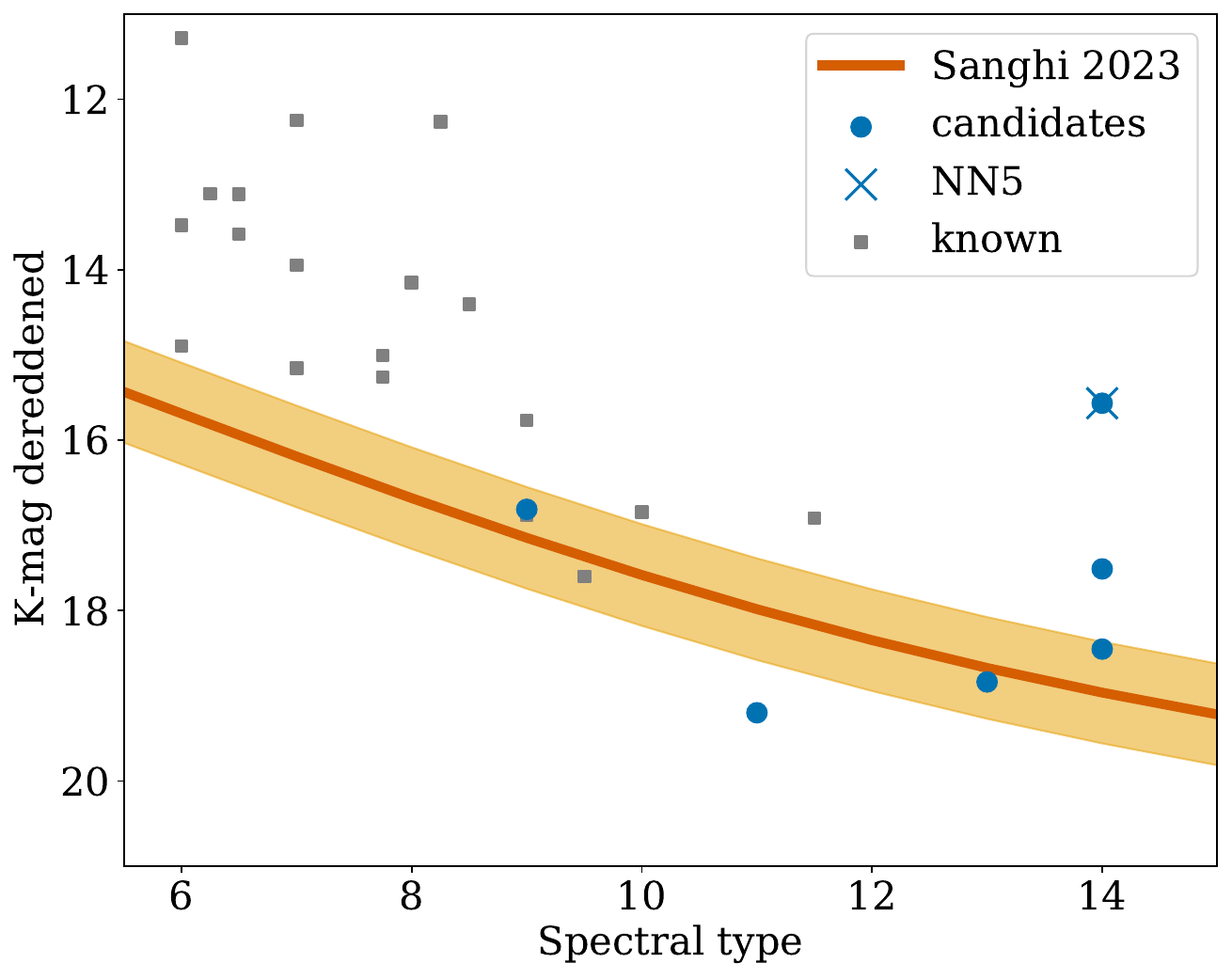} 
\caption{Dereddened K-band photometry vs. spectral type for objects in Table \ref{tab:pmo} after spectral fitting. Typical errorbars would be $\pm 1$ subtype in the x-axis and $\pm 0.3$\,mag in the y-axis. The relationship highlighted in orange is taken from \citet{sanghi2023}. Grey symbols mark the positions of confirmed young brown dwarfs in NGC1333, for comparison. The  object below the relation at L1 spectral type is NN2. The one high above the line at L4 (labelled with a $\times$ symbol) is NN5, which shows clear infrared excess. \label{fig:phot}}
\end{figure}

Figure~\ref{fig:phot} shows an HR diagram of the sample by plotting the spectral type found in the fit against the dereddened K-band magnitude, again using the extinction found in the fit. Dereddening was applied using the extinction law from \citet{wang2019}. The orange band highlights the spectral type vs. K-band relation from \citet{sanghi2023} for young brown dwarfs, shifted to the distance of NGC1333. This relation has been derived for objects older than 10\,Myr. The grey points are known brown dwarfs; those objects appear above the published relation for slightly older objects -- a result of their larger radii compared to their older siblings. Five of our candidates also appear above or around the relation. Based on their K-band brightness, they are plausible cluster members. One of them (NN2), at spectral type L1, is below the relation by about one magnitude. At face value, NN2 is too faint to be a cluster member at its estimated spectral type. One possible explanation for the underluminosity is an edge-on disk blocking parts of the light. Alternatively, NN2 may be a background object.

It is indeed conceivable that a small number of brown dwarfs in the foreground or background of the cluster could be present in our sample. \citet{scholz2022} estimated this number statistically, specifically for this survey in NGC1333, using the survey footprint and the estimated depth, which is achieved in the observations (as shown in Figure~\ref{fig:magnitudes}). From this calculation, we expect 1.4 contaminating field brown dwarfs in the survey, with a plausible range from 0.5 to 4.0. This comes with a strong spectral type dependence; the contamination is negligible for spectral types later than L5. \citet{scholz2022} also point out that these estimates should be considered upper limits. We conclude that among the objects in the shortlist with estimated L spectral types, one or two could be field brown dwarfs, perhaps the one that appears sub-luminous (see above).

Apart from field brown dwarfs, other potential sources of contamination are reddened background stars, as well as young stellar objects with excess emission and/or embedded in dense material. To guard against these possibilities, we repeated the spectral fits with reddened blackbodies as templates, using temperature and $A_V$ as free parameters. A blackbody fit does not produce sensible results for any of these objects.

One object among our candidates (NN5), classified as L4 at moderate extinction, has been previously detected in the mid-infrared and shows clear excess emission. It has a published magnitude of 14.4 at 3.6\,$\mu$m, 13.6 at 4.5\,$\mu$m, and 13.0 at 5.8\,$\mu$m \citep{getman2017}. This is a clear sign that it is in fact a young member of the cluster. Due to the degeneracies between spectral type, extinction, and infrared excess when fitting spectra, the spectral type and mass for this object requires further confirmation. At an estimated mass of $\sim5\,M_{\mathrm{Jup}}$, NN5 is a good candidate to be the lowest-mass object found in this cluster thus far and the lowest-mass object with a disk in any region identified to date. 

In summary, our survey adds a small number of new late M/early L candidates to the existing census of NGC1333. Given their spectral types and assuming cluster membership, these objects would have masses well below the deuterium burning limit. Taking the adopted spectral types at face value, their temperatures range from 1500--2500\,K \citep{sanghi2023}. According to current theoretical evolutionary tracks \citep{phillips2020}, and assuming ages of 1--3\,Myr, this effective temperature range corresponds to masses of 5--15\,$M_{\mathrm{Jup}}$. Our candidates include some of the lowest mass free-floating objects with spectroscopic information to date. We note that further confirmation of cluster membership and spectral type is needed for these candidates. 

\vspace{1em}
\section{Searching for very low mass binaries}
\label{sec:binaries}

Multiplicity is a key property of young stellar and substellar objects \citep[e.g.][]{ahmic2007,defurio2022}. Previous studies have also identified some binaries in the planetary mass regime \citep{jayawardhana2006, brandeker2006,fontanive2020}. Here we searched for close visual pairs of objects among the point sources seen in our NIRISS images and evaluated their properties spectroscopically.

\subsection{Survey for visual pairs}

We conducted a visual examination of the NIRISS F150W and F200W images to identify potential binaries in the observed fields. We limited the search to sources with separations of $<1$\,arcsec, corresponding to a maximum separation of 300\,au for members of NGC1333. Six visual pairs were identified by eye, shown in Figure~\ref{fig:all_binaries} with details listed in Table~\ref{tab:binaries}, and named sequentially from NN7--NN12. 
%When referring to only one object from each pair in future discussion, we label the primary (brighter) object as NNX-A (or fully, \mbox{NIRISS-NGC1333-X-A}), and the companion (fainter) object as NNX-B, where ``X'' is the corresponding index number.

All six pairs were originally included in the full catalog (see Section~\ref{sec:source_catalog}). However, all except for NN8 were categorized as single sources, thus causing both spectra to be combined together during the spectral extraction process. For these, we attempted to separate and extract spectra for each member of the pair independently.

First, we created a modified version of the source catalog that contained only the rows corresponding to the binary candidates. Each row was duplicated (assigning a unique ID to the duplicated source), except for NN8 for which both components were already listed separately. This new catalog was then used as an input to Stage 2 of the spectral extraction process described in Section~\ref{sec:spectral_extraction}, resulting in \textit{\_cal.fits} data products containing two identical slope image cutouts for each binary candidate (except NN8, as mentioned). Following this, the cutouts for a given binary pair were modified manually: in the first cutout, the rows containing the spectral trail of the fainter companion were set to NaN, and vice versa for the brighter object in the second cutout. For binaries NN7 and NN12, the spectral trails of the primary and secondary sources overlapped since the objects are aligned along the same direction as the dispersion from the GR150C grism, which prevents their separation.

The background subtraction algorithm was then applied to the modified binary cutouts, before continuing with Stage~3 processing. Finally, we carried out the same spectral fitting process as described in Section~\ref{sec:spectral_fitting} for each individual spectrum, resulting in a spectral type and an estimate for the optical extinction.

\begingroup

\begin{figure*}[p]
    \centering
    \includegraphics[width=\textwidth]{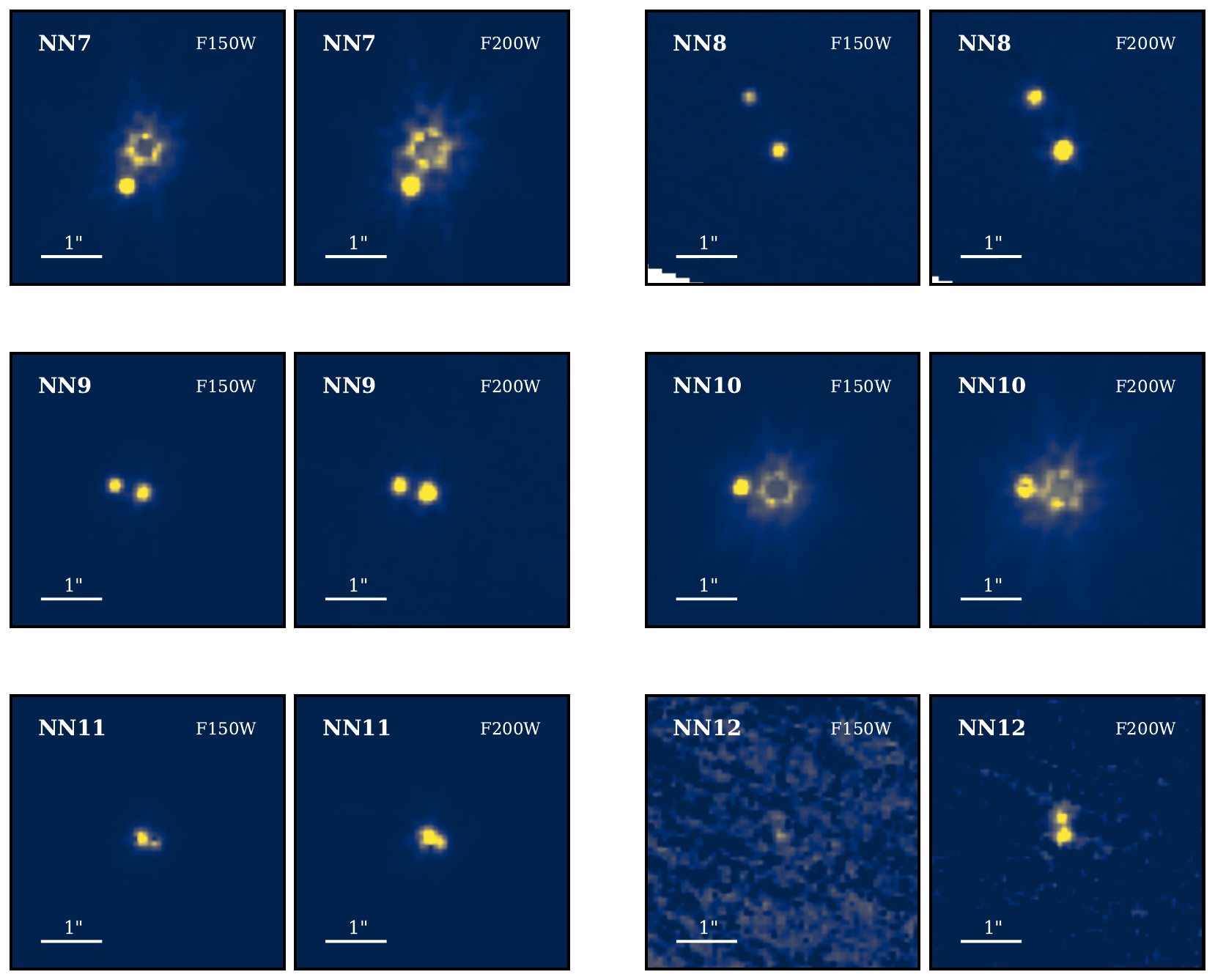}
    \caption{Six binary system candidates identified in the NIRISS F150W and F200W images of NGC1333 with a separation of $<1$\,$''$. Each pair are shown in the world coordinate system, with the F150W image on the left and the F200W image on the right. The color scales are normalized differently for each pair to clearly see the companion objects, thus the difference in brightness between the images for two pairs is not representative of their comparative flux. The primary objects for binaries NN7 and NN10 are saturated in the images. See Table~\ref{tab:binaries} for measured coordinates and separations.}
    \label{fig:all_binaries}
\end{figure*}

\begin{table*}[b]
\centering
\caption{List of binary pairs in NGC1333 with a separation of $<1$\,$''$. The systems are labelled NIRISS-NGC1333 (abbreviated NN) followed by an index that continues from Table~\ref{tab:pmo}. Individual objects are referred to with subscripts A and B, with A being the brighter source. Spectral types are the best fits as described in Section~\ref{sec:spectral_fitting}. The listed K-band magnitudes correspond to the unresolved pairs, taken from UKIDSS-GPS \citep{lucas2008}, except for NN12 where we take it from our own photometry.}
\begin{tabular*}{\textwidth}{c@{\extracolsep{\fill}}cccccccc}
\toprule
ID & RA$_{\rm{A}}$ ($^\circ$)     & Dec$_{\rm{A}}$ ($^\circ$)    & RA$_{\rm{B}}$ ($^\circ$)     & Dec$_{\rm{B}}$ ($^\circ$)    & Separation ($''$)  & SpT$_{\rm{A}}$ & SpT$_{\rm{B}}$ & K (mag)\\ % &Separation (au) \\ 
\midrule                   
% Original IDs
NN7  & 52.217171 & 31.274817 & 52.217259 & 31.274674 & 0.582 & --  & --  & 13.85 \\ %& 174.516 \\ % ID 3, B3
NN8  & 52.241157 & 31.318088 & 52.241284 & 31.318303 & 0.866 & M1  & M1  & 17.20  \\ %& 259.763 \\ % ID 175/177, B4
NN9 & 52.250120 & 31.287049 & 52.250252 & 31.287076 & 0.418 & M1  & M1  & 16.13 \\ %& 125.33  \\ % ID 46, B2
NN10 & 52.264167 & 31.311012 & 52.264342 & 31.311039 & 0.547 & M8  & M9  & 13.90 \\ %& 164.075 \\ % ID 135, B1
NN11 & 52.264840 & 31.298110 & 52.264791 & 31.298085 & 0.177 & M0  & M8  & 16.10 \\ %& 53.129  \\ % ID 85, B5
NN12 & 52.279337 & 31.401251 & 52.279349 & 31.401321 & 0.255 & --  & --  & 20.7    \\ %& 75.6    \\ % ID 516, B6
\bottomrule
\end{tabular*}
\label{tab:binaries}
\end{table*}

\endgroup

\subsection{Results of the binary search}
\label{sec:binaries_results}

With the method outlined above, we find six visual pairs with separations of $<1$\,arcsec. As mentioned above, for two of those (NN7 and NN12), it is not possible to extract separate spectra, which means we cannot report spectroscopic information. While NN7 appears as a bright star based on the combined spectrum, NN12 is very faint and the combined spectrum therefore is inconclusive.

Two more objects, NN8 and NN9, have early M-type components. For NN9, a previous ground-based spectrum was reported by \citet{scholz2012b}; they conclude that this is not a substellar member of the cluster, but may still be a young star. In addition, NN11 was found to have an early M primary star and late M secondary -- if this system is part of the cluster, it would consist of a substellar companion and a low-mass star. The components of NN11 are separated by only 0.177\,$''$ (or $\sim53$\,au if they are members of NGC1333). We note that the two spectra were only visibly separated by one pixel in the slope image cutouts, therefore, these objects are at the limit of how close we can spectroscopically resolve two targets using this JWST/NIRISS WFSS instrument setup without further analysis steps.

The most interesting pair found in our survey is NN10. The primary is a previously known brown dwarf \citep[\mbox{SONYC-NGC1333-8:}][]{scholz2012a}, reported with a spectral type of M7-M8 in the literature. The combined source of NN10 is known to exhibit infrared excess emission, interpreted as evidence for a disk \citep{gutermuth2008}, which also indicates youth consistent with cluster membership. The NIRISS F150W and F200W images reveal for the first time that this source is in fact a binary, with a separation of 0.547\,$''$ (corresponding to $\sim164$\,au, assuming the objects are members of NGC1333 at a distance of 300\,pc). To our knowledge, this is the first wide binary identified by spectroscopy among the known very low mass/substellar members in NGC1333. Our spectral fitting yields spectral types of M8 for the primary and M9 for the secondary; thus, NN10 is likely a brown dwarf with a wide planetary-mass companion, and thus an important benchmark in the discussion of substellar formation paths (see Section~\ref{sec:discussion}). The NIRISS WFSS spectra and template fitting results for both objects in NN10 are shown in Figures~\ref{fig:resultsNN10A} and \ref{fig:resultsNN10B} in Appendix~\ref{sec:appendix_figures_binaries}.
We note that the previously reported binary ASR\,28, with an unconfirmed planetary mass companion to a brown dwarf \citep{greissl2007}, also has infrared excess \citep{gutermuth2008}, but is not covered by our field.

Given the typical density of point sources in our fields, it is possible that the identified binaries are a result of chance alignment. For our field area ($2.2'\times 2.2' \times 7$), and the number of compact sources in our catalog (585), the density of point sources is 0.0044 per arcsec$^2$. We searched for companions within 1\,arcsec, and can realistically detect objects down to 0.2\,arcsec, which means the search area is 3\,arcsec$^2$. Thus, the likelihood that any of the compact sources in the field has an object within that radius by chance alignment is 1.3\,\%. For the 585 objects in our catalog, this translates to about eight coincidental companions; i.e. in principle, all of the companions reported here could be chance projections. However, with one exception all companions we do find have separations of 0.6\,arcsec or less (i.e within a search area of 1\,arcsec$^2$). For chance projections we would expect two thirds  of the candidates to be at separations 0.6--1.0\,arcsec. This points to a high chance some of the pairs we identify are physical binaries. In addition, at least one of them has clear evidence of youth in the form of infrared excess, as discussed above. 

Based on our search with the parameters defined above, the binary fraction among the known brown dwarfs (spectral types M6--M9) in our field is 1/19 (or $\sim5$\,\%), for separations of 50--300\,au. For most of this separation range we are able to detect objects down to the photometric completeness limit (as demonstrated by NN12), corresponding to masses comparable to that of Jupiter. This substellar binary fraction is consistent with previous estimates in other regions \citep{ahmic2007}, albeit affected by low number statistics. Our result also complements a ground-based search for companions to brown dwarfs in this cluster \citep{scholz2009}, which found no companions for separations between 300--1000\,au. For planetary mass objects with L spectral type, the fraction is only weakly constrained, as 0/11, or an upper limit of $<10$\,\%. This ratio relies on the fact that we have 11 objects with spectral type M9 or later in our catalogue (Section \ref{sec:new_candidates}). We do not find spectroscopically confirmed ``Jupiter Mass Binary Objects'' (JuMBOs, i.e. pairs of planetary-mass objects), reported by \citet{pearson2023} in their ONC survey. The possible exception is the faint pair NN12, which, if confirmed as a cluster member, would be a JuMBO in NGC1333 with a total mass of a few Jupiters.

\section{Discussion: FFPMOs and their formation}
\label{sec:discussion}

\subsection{The fraction of FFPMOs in NGC1333}

The NIRISS WFSS survey presented in this paper revealed a small population of objects with properties consistent with cluster membership, and estimated masses between $\sim 5$ and 15\,$M_{\mathrm{Jup}}$. Specifically, we find six objects with spectral types of M9 to mid L that are plausible young free-floating planetary-mass objects in this cluster. Two of those could be in the background, and some may turn out to have earlier spectral types and higher extinctions. Importantly, none of the candidates show a possible signature of methane absorption that is characteristic for T-dwarfs, pointing to a paucity of those objects in NGC1333. Adding in the $\sim 20$ known objects with spectral type M9 or later \citep{scholz2023}, the current tally for FFPMOs in this cluster stands at around 20--30. Thus, FFPMOs appear to constitute a small fraction of the total cluster population. Assuming a total population of $\sim 200$ stars and brown dwarfs \citep{luhman2016}, 20--30 objects correspond to 10--15\,\%. Given the uncertainties in converting from spectral type to mass (see Section \ref{sec:instrument_setup}), some of these objects may actually have masses above the deuterium burning limit, therefore, we consider it plausible that the actual fraction may be as low as $\sim5$\,\%.

\subsection{Comparison with other regions}

In a ground-based photometric study, \citet{miretroig2022} found 70--170 planetary-mass objects in the widespread OB association Upper Scorpius and the star forming region $\rho$ Ophiuchus. A subset of them were spectroscopically confirmed, pointing to negligible contamination rates \citep{bouy2022}. The estimated masses of this population are between 4 and 13$\,M_{\mathrm{Jup}}$. The fraction of free-floating planetary-mass objects, relative to stars and brown dwarfs, was estimated as 2--7\,\%. This is somewhat lower, but given the differences in sensitivity, it is still consistent with our result in NGC1333. Upper Scorpius is a region rich in OB stars, whereas NGC1333 is not. The similarity of the FFPMO fractions in the two regions implies that the presence of massive stars alone cannot be a significant factor in the formation of FFPMOs.

A recently published paper by \citet{luhman2024} presented first results from a combined \mbox{NIRCam} and NIRSpec survey of the central portion of IC348, the sibling to NGC1333, located in the same star forming complex. IC348 has about twice the number of cluster members as NGC1333, but its stellar density is lower \citep{scholz2013}. It is also often thought to be slightly older than NGC1333. \citet{luhman2024} identify three possible planetary-mass objects, adding to the 22 previously known \citep{luhman2016}. Assuming $\sim500$ cluster members in IC348, this corresponds to a \mbox{FFPMO} fraction of approximately 5\,\%, at the low end of the plausible range we found here for NGC1333. The survey by \citet{luhman2024} covers only a small part of IC348, thus the census is likely incomplete.

A third region with comparable survey data is the Orion Nebula Cluster (ONC). Based on preliminary results using JWST/NIRCam, reported by \citet{pearson2023} in a preprint, the number of planetary-mass objects in the ONC might be as high as $\sim 500$. If their results hold, there does not seem to be a drop-off in numbers below 5\,$M_{\mathrm{Jup}}$, and \mbox{FFPMOs} would constitute $\sim 25$\,\% of the $\sim 2000$ cluster members. The completeness of the survey is not characterized yet. For comparison, the most recent HST survey in the ONC \citep{robberto2020,gennaro2020} found about 200 FFPMOs down to masses of a few Jupiters, corresponding to 10\,\% of the cluster membership. The planetary-mass candidates from these surveys are still awaiting confirmation; only a small number of planetary-mass objects in the ONC have been confirmed with spectra \citep{weights2009}. If the number of \mbox{FFPMOs} in the ONC ends up being closer to 500, it would indicate an excess of FFPMOs compared to other regions, including NGC1333. The most obvious difference between the ONC and the other nearby star forming regions is its high stellar density -- twice that of NGC1333 and IC348, and even higher in the center \citep{muzic2019}. This may be a factor in the formation of FFPMOs, as further discussed below.

One of the major problems of the surveys cited above, including ours, is the lack of robust, comparable information on completeness and depth, which hampers comparisons and should be a focus of future studies. A further problem is that in some star forming regions, the age and age dispersion is poorly characterised; therefore it is possible that a few objects classified as FFPMOs based on spectral type may in fact have masses above the deuterium burning threshold.

\subsection{Comparison with simulations}

In the mass domain discussed in this paper, we expect to find two populations: the lowest mass objects that formed like stars, and ejected giant planets (see Section~\ref{sec:intro}). \citet{scholz2022} estimate the expected numbers for different JWST programs, based on prior knowledge about the stellar initial mass function (IMF), the observed rate of giant planets, and predictions from published simulations. For three of these programs (targeting the ONC, NGC1333, and IC348) we now have the first results. \citet{scholz2022} find that the fraction of ejected and now free-floating giant planets (relative to the total number of stars and brown dwarfs in a cluster) is expected to be 1--5\,\%, while the fraction of objects from a log-normal stellar mass function should be only 0.25\,\%. 

The numbers found so far in NGC1333 slightly exceed what is expected for ejected, free-floating planets, and far exceed what is expected from a log-normal stellar mass function. The latter finding has been noted before in other regions \citep{miretroig2022,muzic2019}. Also, for objects formed like stars, we expect a decline of the numbers going to lower masses. Conversely, ejected giant planets should not show such a decline \citep{vanelteren2019}. Thus, the lack of T dwarfs in NGC1333 might tell us that we are observing the tail-end of the stellar mass function, a fundamental limit of star formation. There may still be some ejected planets in this cluster, but not in significant numbers. A similar conclusion was reached by \citet{parker2023}, by comparing the spatial distribution of FFPMOs in NGC1333 with that expected for free-floating planets. 

One of the main mechanisms for planet ejections is close encounters with other stars in a clustered environment. Several papers specifically point out that the rate of ejection should depend strongly on the initial stellar density in the cluster \citep{parker2012,daffernpowell2022}. This could lead to a strong environmental dependence in the formation of free-floating planets. The paucity of T dwarfs in NGC1333 may be a result of an insufficient stellar density to facilitate ejections. It is conceivable that other clusters with higher density (specifically, the ONC) eject planets at a higher rate.

As pointed out above, the discussion here is based on early results, which will need to be confirmed in the coming years. Moreover, the production of free-floating planets is also expected to be affected by a range of other factors -- for example, it should be highly sensitive to the age of the cluster \citep{vanelteren2019}. Exploring the interplay of the relevant conditions will require observations of diverse environments, beyond those studied thus far.

\vspace{-0.2em}
\section{Summary}
\label{sec:summary}

We present first results from a deep spectroscopic survey of the young cluster NGC1333, conducted with JWST and NIRISS in WFSS mode. We observed seven WFSS fields in the central parts of the cluster, covering a substantial fraction of the known population. Analyzing our images and spectra, we show that the design goals of the survey have been mostly achieved. In particular, we reach a photometric completeness at $K\sim 21$ with many objects at fainter magnitudes. The spectroscopy yields useful spectra for objects down to $K\sim 20.5$. 

Overall, we identify approximately 600 compact sources in our survey area, out of which 114 have plausible stellar or substellar spectra. Those spectra were analyzed by comparing with templates for young M-, L- and \mbox{T-dwarfs}. Most are classified as early or mid M-dwarfs. We re-identify 19 known substellar members in this cluster and show that our spectral typing works with uncertainties of one subtype for mid M- to early L-dwarfs. 

Our sample includes six objects with spectra that are best matched by spectral templates with type of M9 to L4. While some of these are known photometrically in the literature, we present their spectral classification for the first time. One has clear infrared excess indicative of a dusty disk. While all six objects require additional confirmation, they are good candidates to be planetary-mass members of NGC1333, with estimated masses of approximately \mbox{5--15\,$M_{\mathrm{Jup}}$}. Notably, none of the objects with spectra shows clear signs of methane absorption characteristic for T-dwarfs (or masses below 4\,$M_{\mathrm{Jup}}$), despite the fact that our survey reaches the required flux levels. 

Seen in context, NGC1333 harbors a relatively sparse population of FFPMOs, with a frequency of about 10\,\%, with a current low-mass cutoff at $\sim 5\,M_{\mathrm{Jup}}$. This is consistent with the survey results in Upper Scorpius and IC348, but not with the recent report of an excess of \mbox{FFPMOs} in the ONC using JWST data. An enhancement of free-floating planets in high-density clusters (such as the ONC) is expected if they are formed primarily as a result of stellar encounters with nascent planetary systems.

We also surveyed our images for wide pairs that could be potential very low mass binaries. Our search is sensitive to separations of 50--300\,au at the distance of this cluster. When possible, we report separate spectral classifications for these sources. We resolve one known brown dwarf into two objects; the companion has an M9 spectral type, corresponding to a mass close to the deuterium burning limit.

We demonstrate that slitless spectroscopy with the WFSS mode of NIRISS can be a valuable tool to identify substellar objects in clusters well into the planetary mass regime. In particular, it is an effective method to capture a snapshot of the population. Follow-up studies are, however, needed to characterize candidates more comprehensively. The technique works well when combined with multi-band photometry. It will be most efficient for compact (yet not crowded) clusters with a significant number of brown dwarfs and low or moderate extinction. For a distance of 300\,pc and an age of 1--3\,Myr (like NGC1333), it is possible to achieve a sensitivity limit of 1\,$M_{\mathrm{Jup}}$. This limit will be at higher masses for more distant or older clusters. Thus, NGC1333 may be one of the most suitable regions for this approach.

\vspace{0.5em}
\begin{acknowledgments}
We thank the anonymous referee for a constructive and concise report that helped to improve this paper.
A.S. acknowledges support from the UKRI Science and Technology Facilities Council through grant ST/Y001419/1/. 
K.M. acknowledges support from the Fundação para a Ciência e a Tecnologia (FCT) through the CEEC-individual contract 2022.03809.CEECIND and research grants UIDB/04434/2020 and UIDP/04434/2020. 
R.J. acknowledges support from a Rockefeller Foundation Bellagio Residency. 
M.D.F. is supported by an NSF Astronomy and Astrophysics Postdoctoral Fellowship under award AST-2303911.
D.J. is supported by NRC Canada and by an NSERC Discovery Grant.
\newline
\end{acknowledgments}

\vspace{-0.5em}
\facilities{JWST}

\software{}
Astropy \citep{astropy2013,astropy2018,astropy2022}, 
Numpy \citep{harris2020}, 
Scipy \citep{virtanen2020}, 
Matplotlib \citep{hunter2007}, 
JWST Science Calibration Pipeline \citep{bushouse2023},
Pandas \citep{pandas2022}, 
SExtractor \citep{bertin1996}, 
SAOImage DS9 \citep{ds92000},
Topcat \citep{taylor2005}.

\vspace{1em}
\appendix{}
\section{Planetary-mass candidates in NGC1333 from JWST/NIRISS WFSS}
\label{sec:appendix_figures}
\begin{figure*}[p]
    \includegraphics[width=\textwidth]{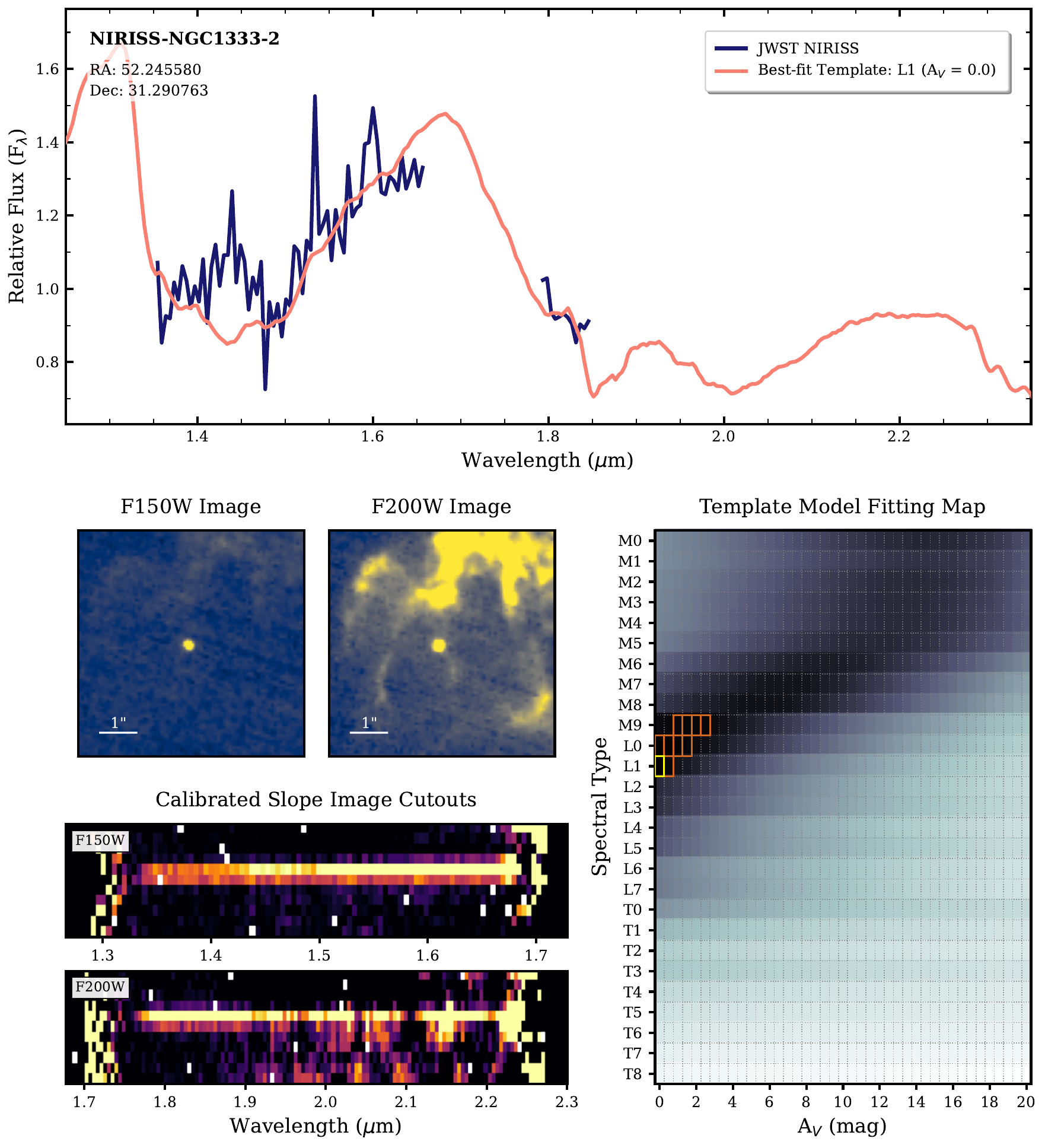}
    \caption{Results of the spectroscopic survey for object NIRISS-NGC1333-2 (NN2). 
    Panels shown are the same as Figure~\ref{fig:resultsNN1}. Most of the F200W spectrum had to be discarded since the spectral trail in the slope image cutout overlapped with a diffuse background region. This is possibly the reason for the excess flux seen around 1.45$\,\mu m$.}
    \label{fig:resultsNN2}
\end{figure*}

\begin{figure*}[p]
    \includegraphics[width=\textwidth]{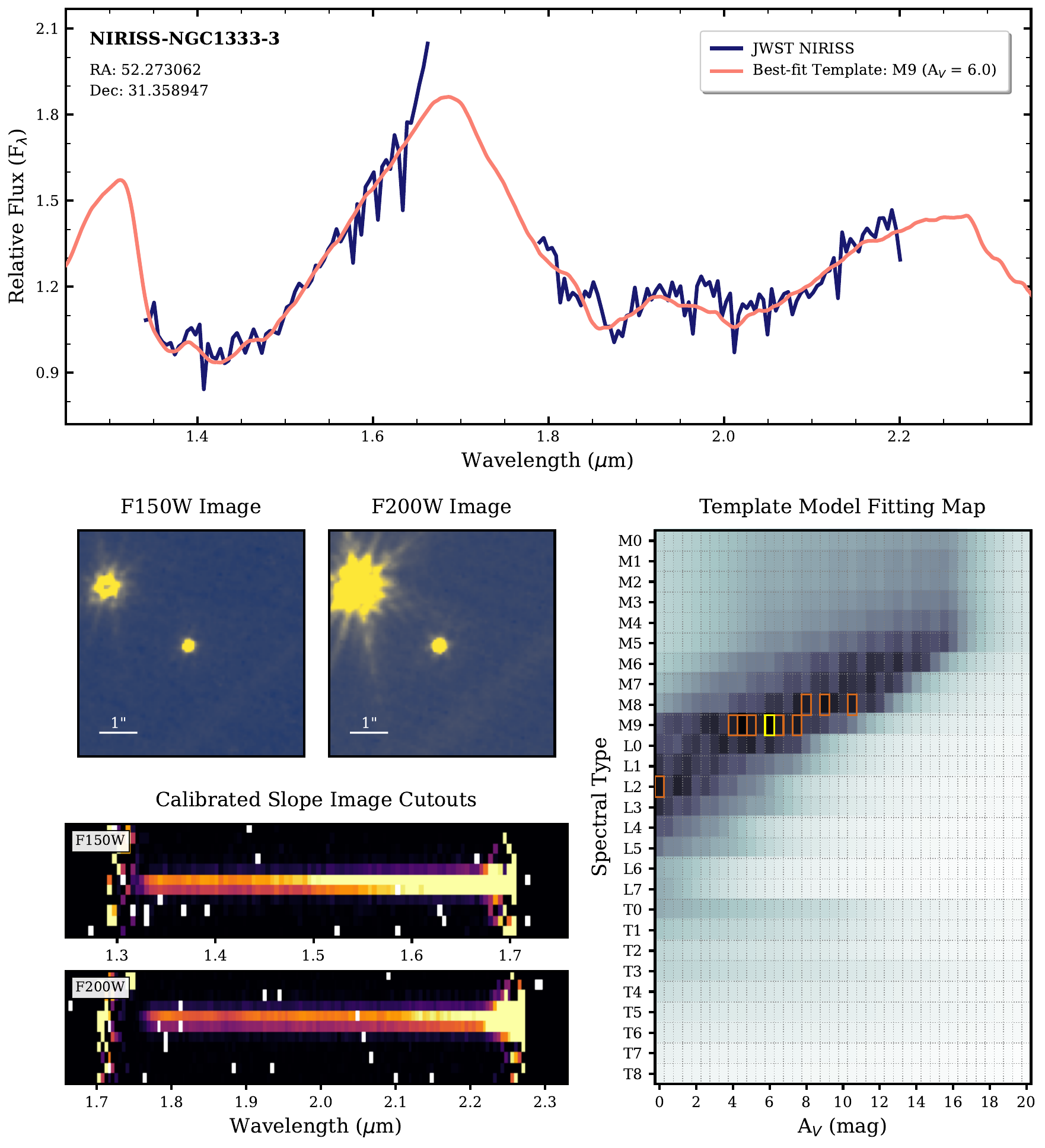}
    \caption{Results of the spectroscopic survey for object NIRISS-NGC1333-3 (NN3). 
    Panels shown are the same as Figure~\ref{fig:resultsNN1}.}
    \label{fig:resultsNN3}
\end{figure*}

\begin{figure*}[p]
    \includegraphics[width=\textwidth]{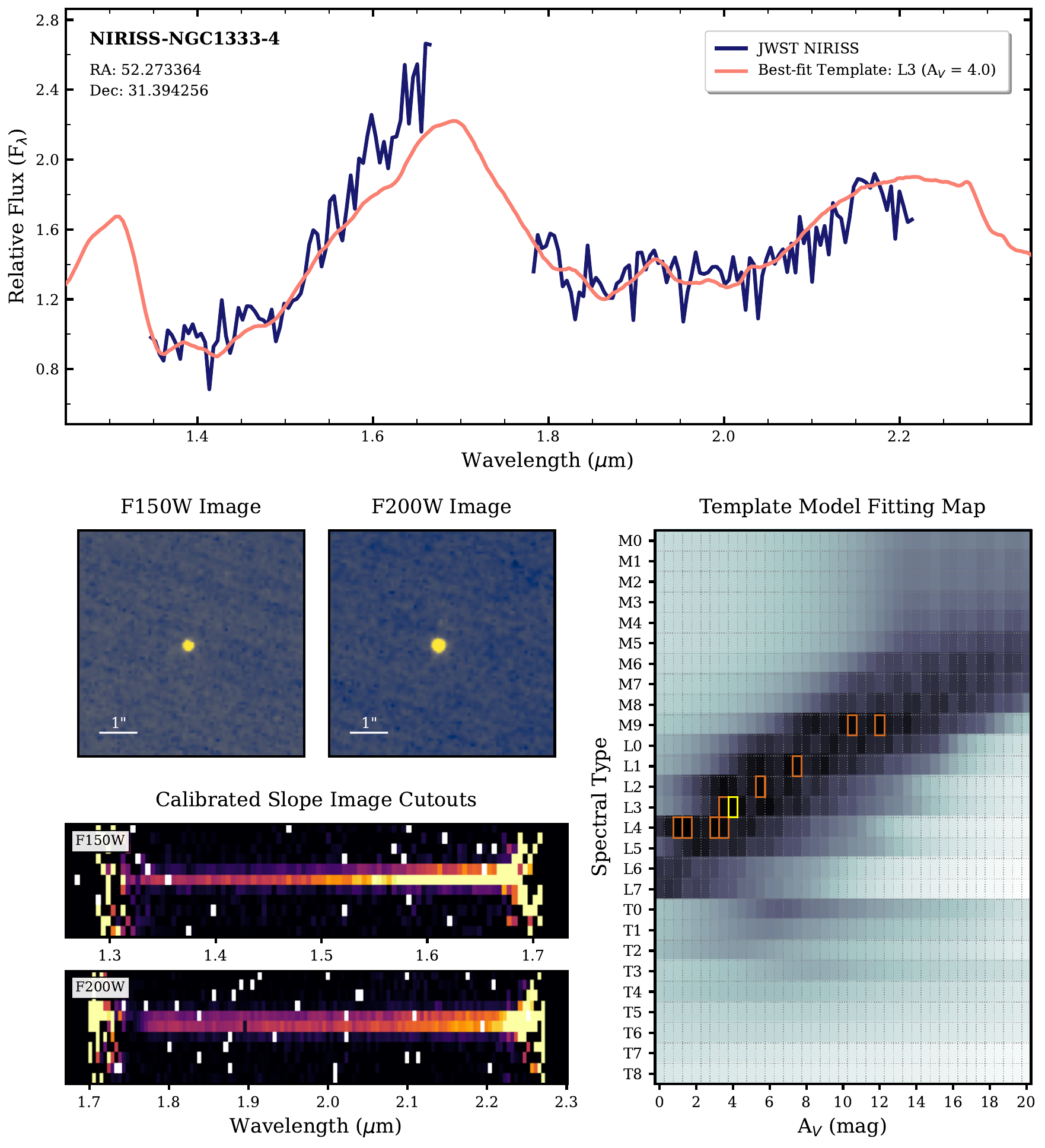}
    \caption{Results of the spectroscopic survey for object NIRISS-NGC1333-4 (NN4). 
    Panels shown are the same as Figure~\ref{fig:resultsNN1}.}
    \label{fig:resultsNN4}
\end{figure*}

\begin{figure*}[p]
    \includegraphics[width=\textwidth]{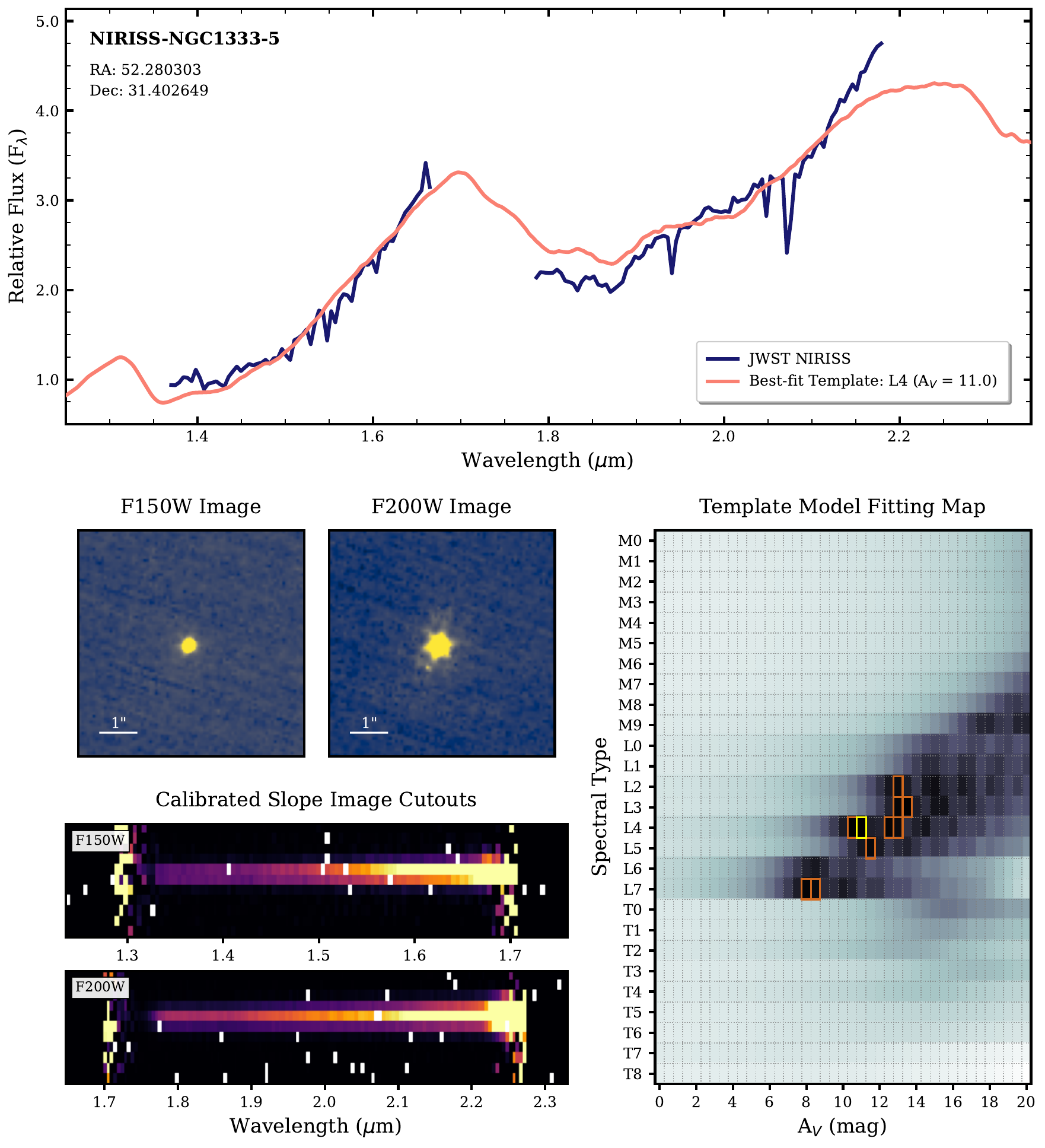}
    \caption{Results of the spectroscopic survey for object NIRISS-NGC1333-5 (NN5). 
    Panels shown are the same as Figure~\ref{fig:resultsNN1}.
    This object has known infrared excess (see Section~\ref{sec:new_candidates}), which may explain the excess flux at $\sim2.2$\,$\mu$m.}
    \label{fig:resultsNN5}
\end{figure*}

\begin{figure*}[p]
    \includegraphics[width=\textwidth]{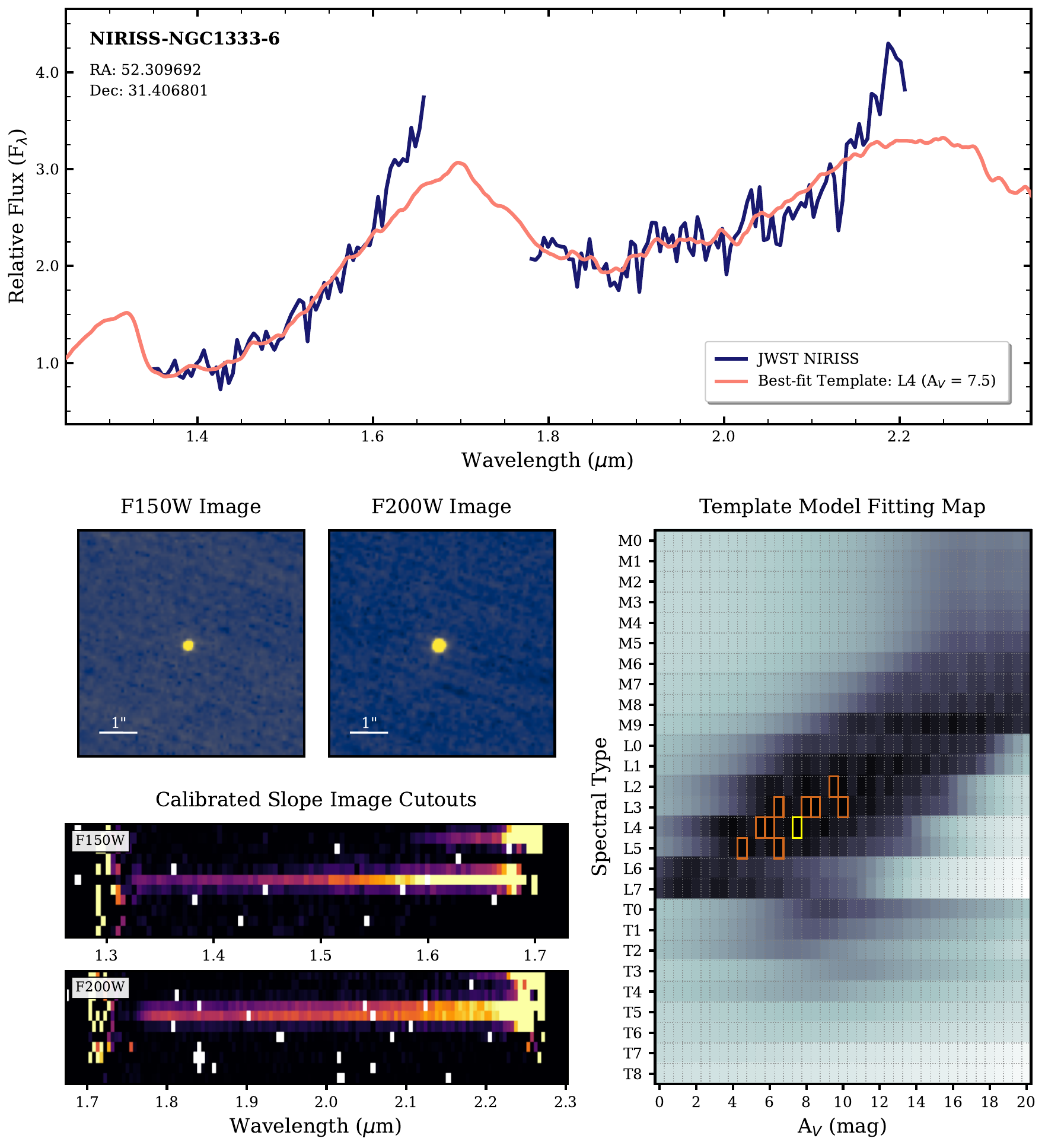}
    \caption{Results of the spectroscopic survey for object NIRISS-NGC1333-6 (NN6). 
    Panels shown are the same as Figure~\ref{fig:resultsNN1}. The calibrated slope image cutouts show that there is contamination from a nearby source (evident by a small trail in the upper right of both cutouts, overlapping the NN6 spectrum at longer wavelengths). This causes excess flux at $\sim1.65$\,$\mu$m and $\sim2.2$\,$\mu$m, and the deviation from the model template in the top panel.}
    \label{fig:resultsNN6}
\end{figure*}

Here we include figures (Figures~\ref{fig:resultsNN2}--\ref{fig:resultsNN6}) showing the NIRISS WFSS images, spectra, and the results from the template fitting, for all objects listed in Table~\ref{tab:pmo} except NN1 (which is described in detail in Figure~\ref{fig:resultsNN1}).

\vspace{1em}
\section{A planetary-mass candidate in a binary system in NGC1333 from JWST/NIRISS WFSS}
\label{sec:appendix_figures_binaries}
\begin{figure*}[p]
    \includegraphics[width=\textwidth]{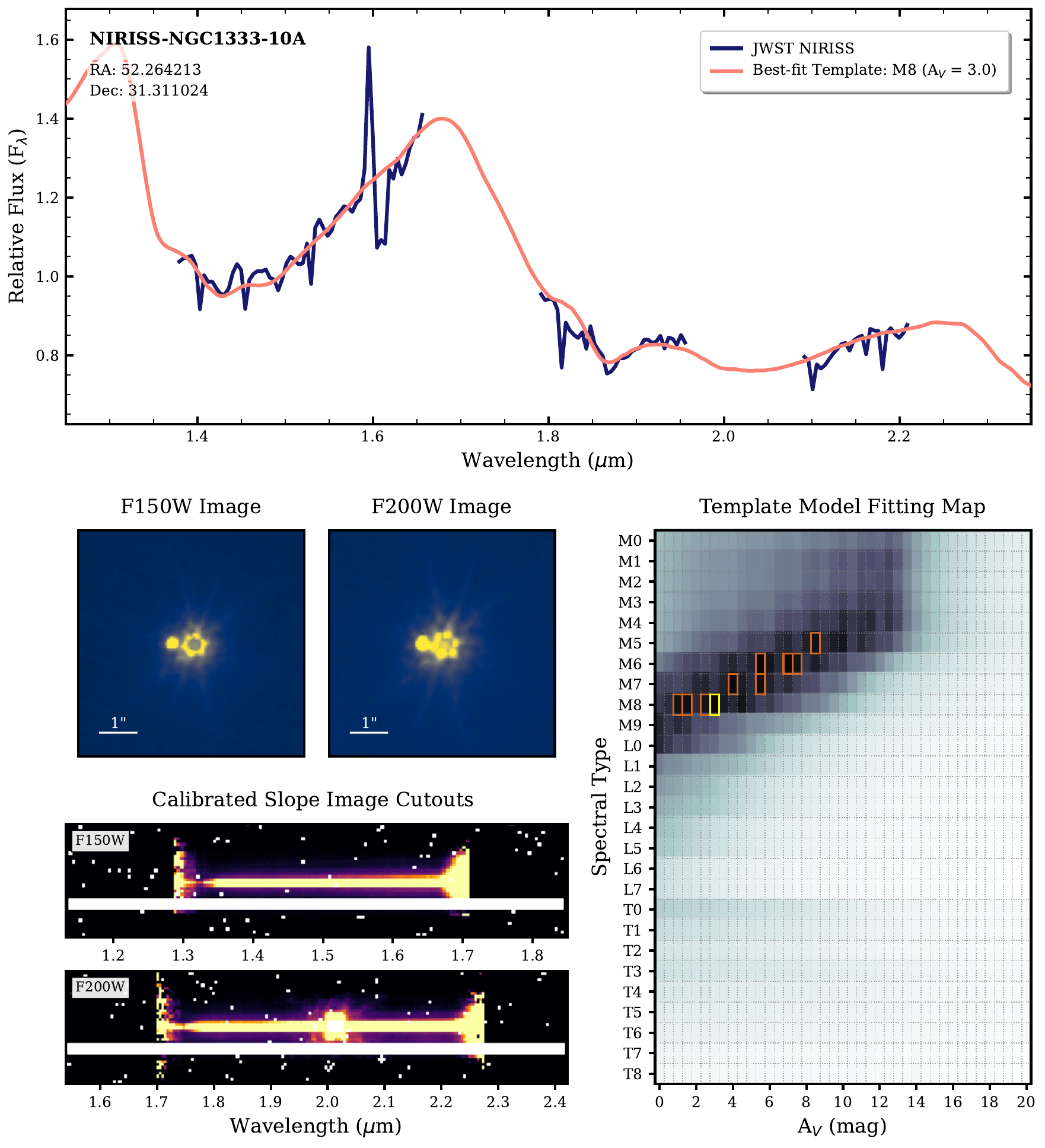}
    \caption{Results of the spectroscopic survey for object NIRISS-NGC1333-10A (NN10A) -- the primary object of the binary NN10 (with NN10B shown in Figure~\ref{fig:resultsNN10B}). 
    Panels shown are the same as Figure~\ref{fig:resultsNN1}. The calibrated slope image cutouts show that there is contamination in the F200W spectrum from a source that overlaps with the spectral trail -- the data in this region is unusable and was removed (evident by the gap in the spectrum in the top panel between $\sim 1.95$--2.1\,$\mu$m). The cutouts also display a large white region (where all values are NaNs) to remove the spectrum of the companion object.}
    \label{fig:resultsNN10A}
\end{figure*}

\begin{figure*}[p]
    \includegraphics[width=\textwidth]{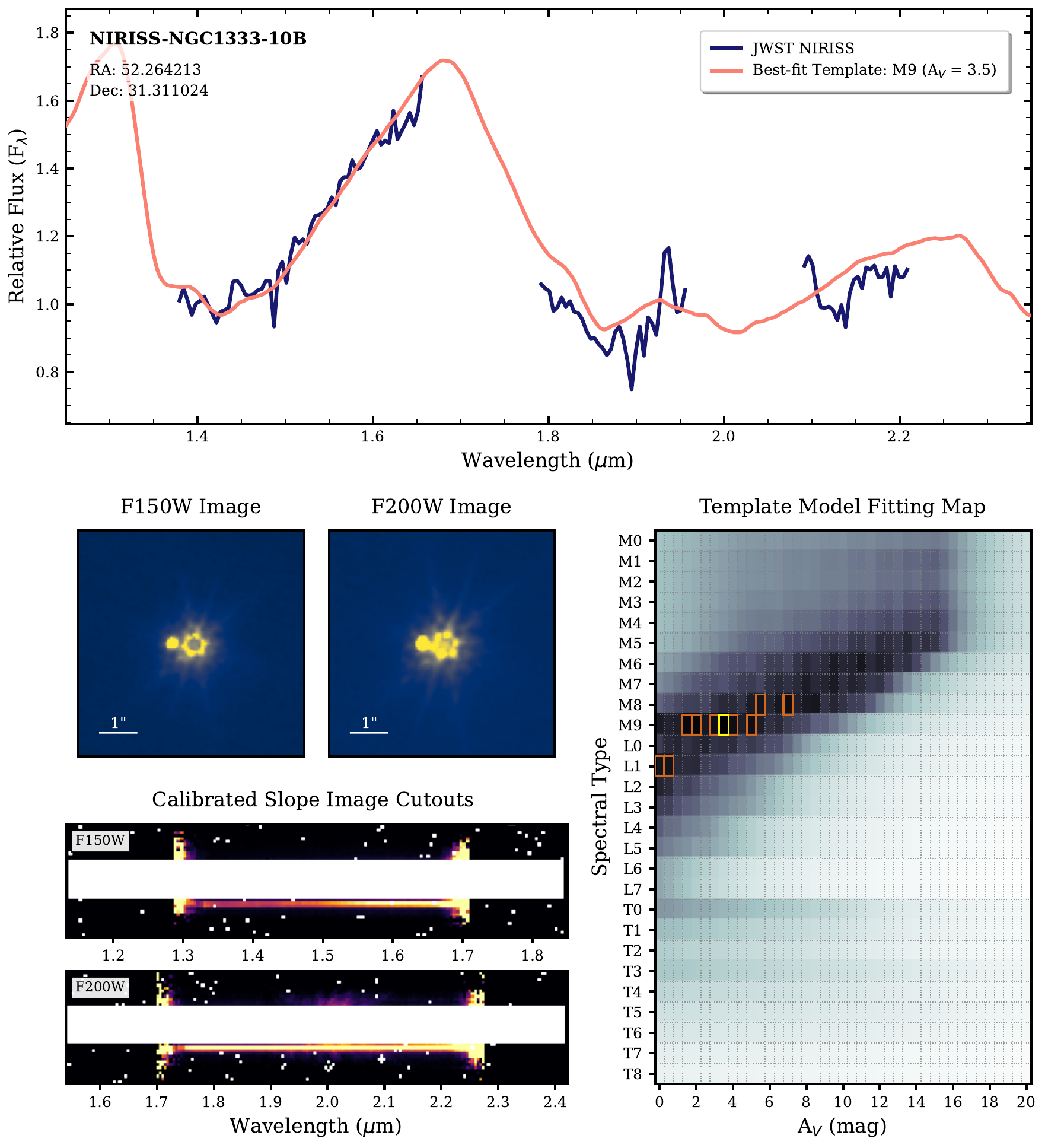}
    \caption{Results of the spectroscopic survey for object NIRISS-NGC1333-10B (NN10B) -- the companion object of the binary NN10 (with NN10A shown in Figure~\ref{fig:resultsNN10A}). 
    Panels shown are the same as Figure~\ref{fig:resultsNN1}. Similarly to NN10A (Figure~\ref{fig:resultsNN10A}), there is contamination in the F200W spectrum from a source that overlaps with the spectral trail -- the data in this region is unusable and was removed (evident by the gap in the spectrum in the top panel between $\sim 1.95$--2.1\,$\mu$m). The cutouts also display a large white region (where all values are NaNs) to remove the spectrum of the primary object.}
    \label{fig:resultsNN10B}
\end{figure*}

Here we include figures (Figures~\ref{fig:resultsNN10A} and \ref{fig:resultsNN10B}) showing the NIRISS WFSS images, spectra, and the results from the template fitting, for the two objects in the binary system NN10. Further discussion can be found in Section~\ref{sec:binaries_results}.

% ##### References
\vspace{1em}
\bibliography{references}

%\bibliography{}{}
\bibliographystyle{aasjournal}

\end{document}